\documentclass[11pt]{article}
\pdfoutput=1 % if your are submitting a pdflatex (i.e. if you have
\usepackage{jheppub} % for details on the use of the package, please
\usepackage[T1]{fontenc} % if needed

\usepackage{bm}
\usepackage{bigdelim}
\usepackage[table,dvipsnames]{xcolor}

\usepackage{hhline}

\def\alphas{\alpha_{\rm s}}
\def\Nc{N_{\rm c}}
\def\Nf{N_{\rm f}}
\def\CA{C_{\rm A}}
\def\qhatA{\hat q_{\rm A}}
\def\Re{\operatorname{Re}}

\def\eps{\epsilon}
\def\beps{{\bm\epsilon}}

\def\p{{\bm p}}
\def\q{{\bm q}}
\def\grad{{\bm\nabla}}

\def\ix{{\rm i}}
\def\fx{{\rm f}}
\def\xx{{\rm x}}

\def\yx{{\rm y}}
\def\ybx{{\bar{\rm y}}}
\def\zx{{\rm z}}
\def\PV{\operatorname{PV}}

\def\virtI{{\rm virt\,I}}
\def\virtIc{{\rm virt\,Ic}}
\def\virtIs{{\rm virt\,Is}}
\def\virtIf{{\rm virt\,If}}
\def\virtII{{\rm virt\,II}}
\def\virtIIs{{\rm virt\,IIs}}
\def\virtIIf{{\rm virt\,IIf}}

\def\NLObar{\overline{\rm NLO}}

\title{The LPM effect in sequential bremsstrahlung: incorporation of
  ``instantaneous'' interactions for QCD}

\author[a]{Peter Arnold,}
\author[b,c]{Tyler Gorda,}
\author[d,e]{Shahin Iqbal}

\affiliation[a]{Department of Physics, University of Virginia,
  P.O.\ Box 400714, 
  Charlottesville, VA 22904, U.S.A.}
\affiliation[b]{Technische Universit{\"a}t Darmstadt, Department of
  Physics, 64289 Darmstadt, Germany}
\affiliation[c]{ExtreMe Matter Institute EMMI and Helmholtz Research
  Academy for FAIR, GSI Helmholtzzentrum f\"ur Schwerionenforschung GmbH,
  64291 Darmstadt, Germany}
\affiliation[d]{National Centre for Physics,
  Quaid-i-Azam University Campus, Islamabad, 45320 Pakistan}
\affiliation[e]{Institute of Particle Physics,
    Central China Normal University, Wuhan, 430079, China}

\emailAdd{parnold@virginia.edu}
\emailAdd{tyler.gorda@physik.tu-darmstadt.de}
\emailAdd{smi6nd@virginia.edu}

\abstract{
  The splitting processes of bremsstrahlung and pair production in a medium
  are coherent over large distances in the very high energy limit,
  which leads to a suppression known as the Landau-Pomeranchuk-Migdal
  (LPM) effect.  We continue study of the case when the coherence
  lengths (formation lengths)
  of two consecutive splitting processes overlap,
  avoiding soft-emission approximations.
  Previous work made a ``nearly-complete'' calculation of the effect of
  overlapping formation times on gluonic splittings such as
  $g \to gg \to ggg$ (with simplifying assumptions such as an
  infinite QCD medium and the large-$\Nc$ limit).
  In this paper, we extend those previous rate calculations from
  nearly-complete to complete by including processes involving the
  exchange of longitudinally-polarized gluons.
  In the context of Lightcone Pertubation Theory, used earlier for the
  ``nearly-complete'' calculation,
  such exchanges are instantaneous in lightcone time and have
  their own diagrammatic representation.
}

\begin{document} 
\maketitle
\flushbottom

%%%%%%%%%%%%%%%%%%%%%%%%%%%%%%%%%%%%%%%%%%%%%%%%%%%%%%%%%%%%%%%%%%%%%%%%%%%%%%%

\section{Introduction}

\subsection {Overview}

When passing through matter, high energy particles lose energy by
showering, via the splitting processes of hard bremsstrahlung and pair
production.  At very high energy, the quantum mechanical duration of
each splitting process, known as the formation time, exceeds the mean
free time for collisions with the medium, leading to a significant
reduction in the splitting rate known as the Landau-Pomeranchuk-Migdal
(LPM) effect \cite{LP1,LP2,Migdal}.%
\footnote{
  The papers of Landau and Pomeranchuk \cite{LP1,LP2} are also available in
  English translation \cite{LPenglish}.
}
The generalization of the LPM effect from QED to QCD was originally
carried out by
Baier, Dokshitzer, Mueller, Peigne, and Schiff \cite{BDMPS1,BDMPS2,BDMPS3}
and by Zakharov \cite{Zakharov1,Zakharov2}
(BDMPS-Z).
A long-standing problem in field theory has
been to understand how to implement this effect in cases where
the formation times of two consecutive splittings overlap.
Several authors \cite{Blaizot,Iancu,Wu} have previously analyzed this issue
for QCD at leading-log order, which arises from the limit where one
bremsstrahlung gluon is soft compared to the other very-high energy
partons.
In a series of papers \cite{2brem,seq,dimreg,4point,QEDnf,qedNfstop,qcd},
we and collaborators have worked on a
program to evaluate the effects of
overlapping formation times without leading-log or soft
bremsstrahlung approximations.
Ref.\ \cite{qcd} presented what we called a ``nearly complete'' calculation
of the relevant rates for the effect of overlapping formation times
on both (i) two consecutive gluon splittings $g \to gg \to ggg$ and
(ii) related (and equally important)
virtual corrections $g \to gg \to ggg \to gg$ to single splitting $g \to gg$.
The purpose of the present paper is to
turn ``nearly complete'' into ``complete'' (within the context of
the approximations used in earlier work, reviewed below).

\begin {figure}[t]
\begin {center}
  \includegraphics[scale=0.5]{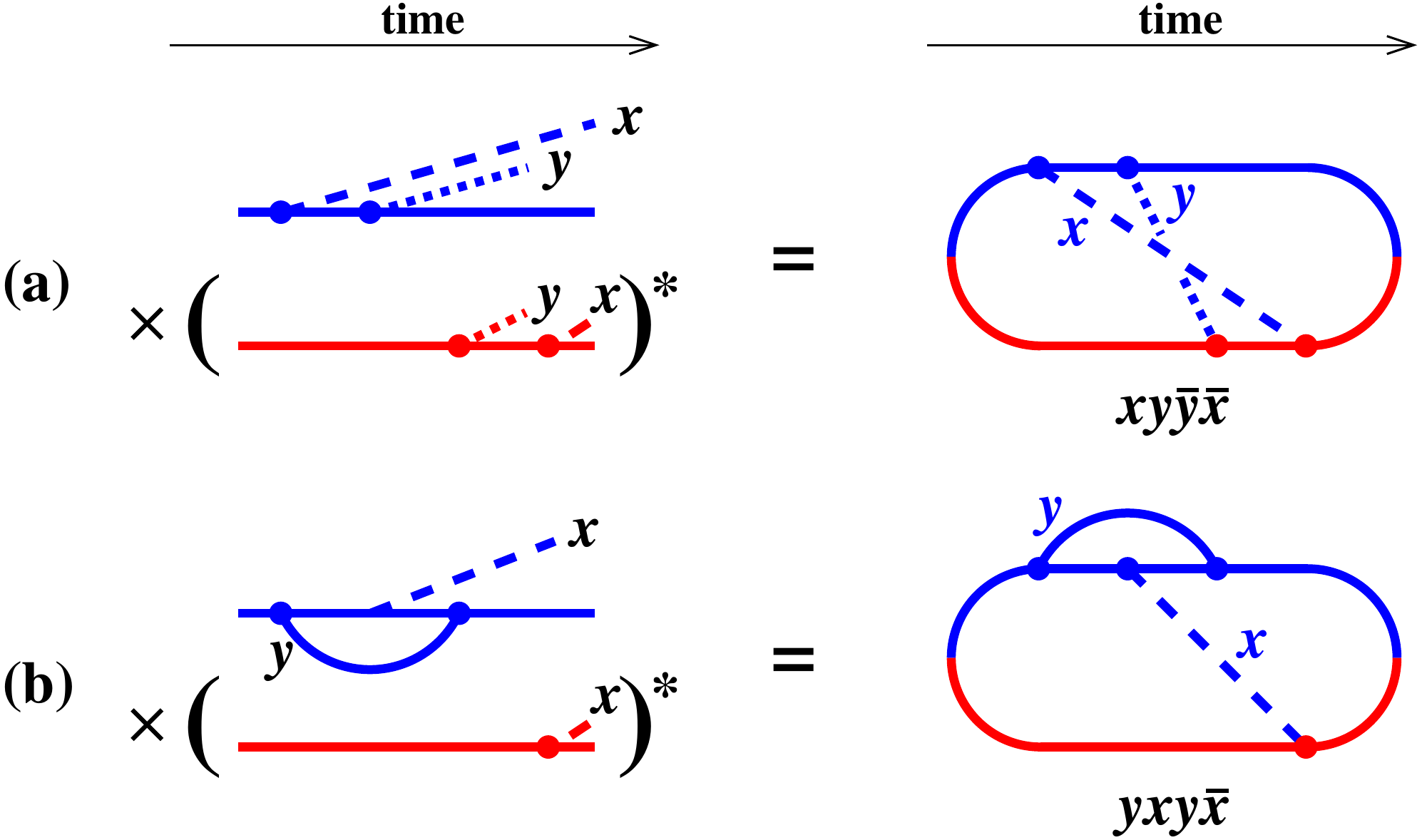}
  \caption{
     \label{fig:previous}
     Two examples (previously computed \cite{2brem,qcd}) of
     interference contributions to the rates for
     (a) double splitting $g {\to} ggg$ and
     (b) virtual corrections to $g {\to} gg$
     (where $y$ is virtual and must be integrated over).
     All lines in these diagrams represent high-energy gluons which
     implicitly and repeatedly interact with the medium (not shown).
     The left side above depicts contributions to the rate,
     obtained by multiplying a contribution to the amplitude (blue)
     by a contribution to the conjugate amplitude (red), with
     a particular time-ordering of all the vertices.
     The right side shows a more compact way of diagrammatically
     representing the same interference contributions, which is
     particularly useful for our implementation and extension
     \cite{2brem,qcd} of
     Zakharov's method \cite{Zakharov1,Zakharov2} for
     organizing and computing the LPM effect.
     In these diagrams, we need not follow a daughter of the splitting
     after its emission has occurred in both the amplitude and
     conjugate amplitude because we will only consider
     $p_\perp$-integrated rates.  (See, for example, section 4.1
     of ref.\ \cite{2brem} for an explicit argument.)
     The (time-ordered)
     diagrams are named $xy\bar y\bar x$ and $xyy\bar x$ here
     according to the convention of refs.\ \cite{2brem,qcd}, summarized
     in the text.
  }
\end {center}
\end {figure}

Fig.\ \ref{fig:previous} shows one example each of time-ordered contributions
to (a) the rate for double splitting $g \to ggg$ with energies
$E \to xE + yE + (1{-}x{-}y)E$ and (b) virtual
corrections (at the same order) to the rate for single splitting $g \to gg$
with energy $E \to xE + (1{-}x)E$.
Each diagram is time-ordered from left to right and
has the following interpretation: The blue (upper)
part of the diagram represents a contribution to the amplitude for
$g \to ggg$ or $g \to gg$, the red (lower) part represents a contribution to the
conjugate amplitude, and the two together represent a particular
contribution to the {\it rate}.  Only high-energy particle lines are
shown explicitly, but each such line is implicitly
summed over an arbitrary number
of interactions with the medium, and the rate is averaged over
the statistical fluctuations of the medium.
See ref.\ \cite{2brem} for details.
The examples shown in fig.\ \ref{fig:previous} are just two of many that
were incorporated into the ``nearly complete'' analysis of rates
in ref.\ \cite{qcd}.  That analysis was carried out in the framework of
time-ordered lightcone perturbation theory (LCPT)
\cite{LB,BL,BPP},%
\footnote{
  For readers not familiar with time-ordered
  LCPT who would like
  the simplest possible example of how it reassuringly
  reproduces the results of
  ordinary Feynman diagram calculations,
  we recommend section
  1.4.1 of Kovchegov and Levin's monograph \cite{KL}.
}
where all the
lines of fig.\ \ref{fig:previous}, for example, represent transverse-polarized
gluons.

\begin {figure}[t]
\begin {center}
  \includegraphics[scale=0.5]{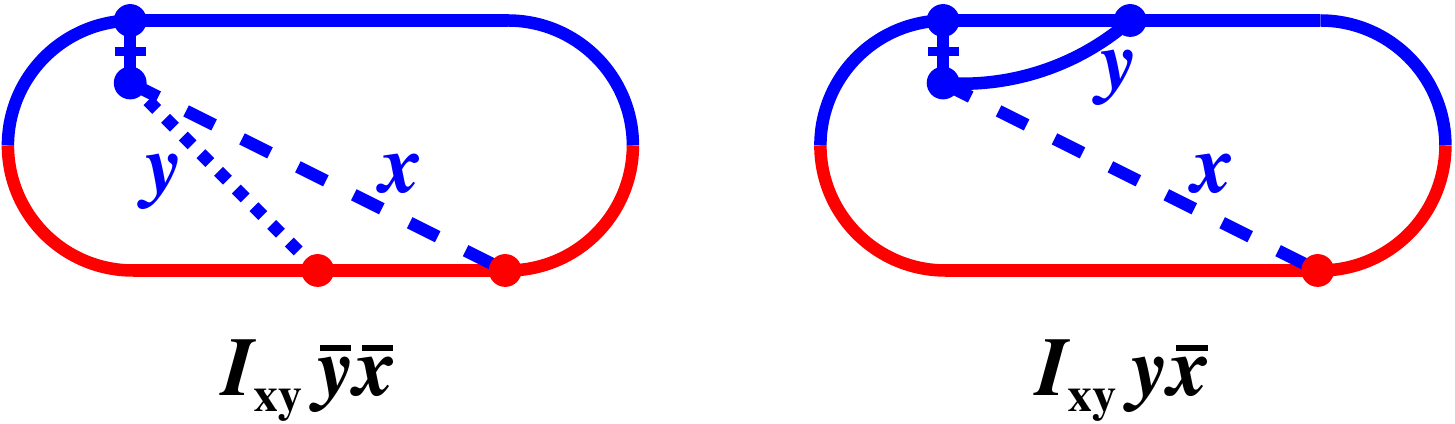}
  \caption{
     \label{fig:examples} 
     Two examples of interference contributions
     involving a
     longitudinally polarized gluon, represented by the vertical line
     crossed by a bar.
     The line is drawn vertically because the interaction is
     instantaneous in (lightcone) time.
  }
\end {center}
\end {figure}

Missing from that analysis were diagrams involving
exchange of a longitudinally-polarized gluon in lightcone gauge.
As we'll review later, such interactions are
instantaneous in lightcone time.
Examples are shown in fig.\ \ref{fig:examples}, where we follow the
standard LCPT convention of using a vertical line (because the interaction
is instantaneous) crossed by a bar to represent
the longitudinally-polarized gluon.  Analogous contributions to overlap
effects in double splitting have previously been analyzed for
large-$\Nf$ QED in ref.\ \cite{QEDnf}, and we will use similar
methods here.

\begin {figure}[t]
\begin {center}
  \includegraphics[scale=0.5]{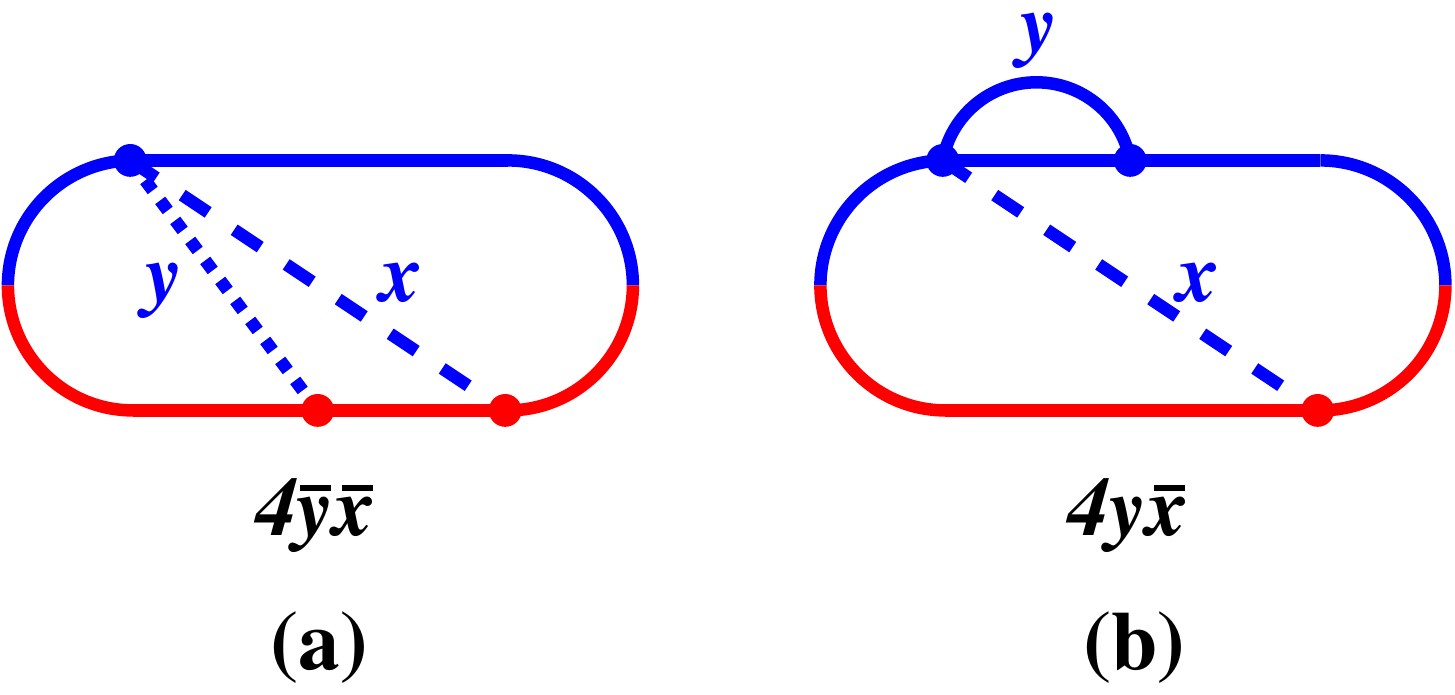}
  \caption{
     \label{fig:4point} 
     Two examples of interference contributions
     involving a fundamental 4-gluon vertex:
     (a) example of double splitting contributions $g {\to}ggg$ calculated in
     ref.\ \cite{4point}, and (b) example of a corresponding virtual
     diagram to be computed in this paper.
  }
\end {center}
\end {figure}

Also missing from the ``nearly complete'' calculation of ref.\ \cite{qcd}
were processes involving the fundamental 4-gluon interactions of QCD,
examples of which are shown in fig.\ \ref{fig:4point}.  Ref.\ \cite{4point}
previously computed such processes in the case of
real double splitting $g {\to} ggg$, such as fig.\ \ref{fig:4point}a,
but the corresponding virtual diagrams, such as fig.\ \ref{fig:4point}b,
have not previously been calculated.

\begin {figure}[t]
\begin {center}
  \includegraphics[scale=0.5]{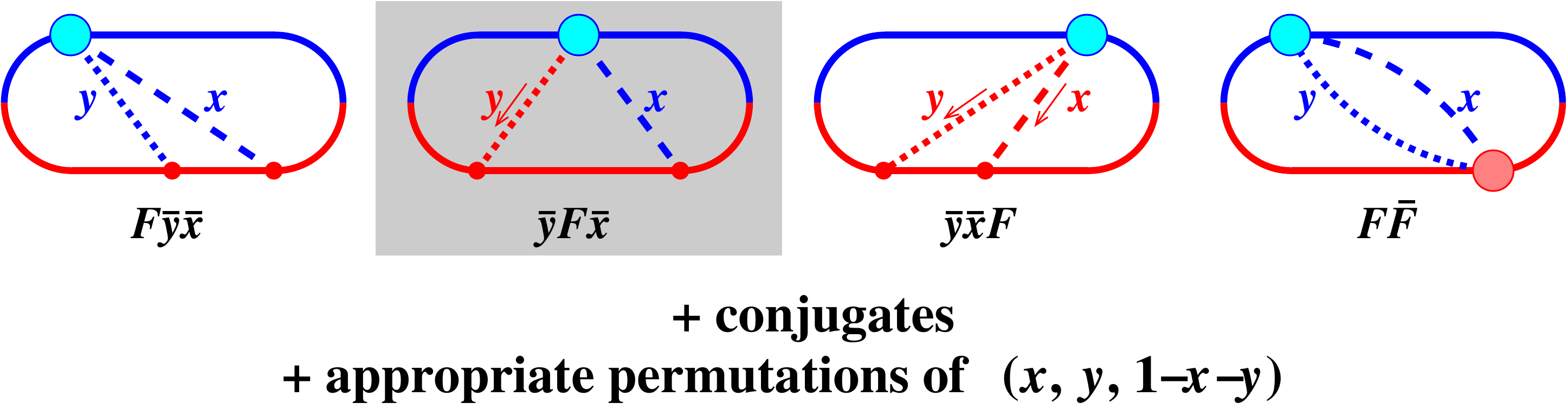}
  \caption{
     \label{fig:FdiagsReal}
     Diagrams for real double splitting $g {\to} ggg$ that involve
     an instantaneous exchange or fundamental 4-gluon interaction.
     See fig.\ \ref{fig:Fvertex} for the meaning of the large circular
     blob.  The diagram drawn on a gray background turns out to be
     exactly zero.
  }
\end {center}
\end {figure}

\begin {figure}[t]
\begin {center}
  \includegraphics[scale=0.5]{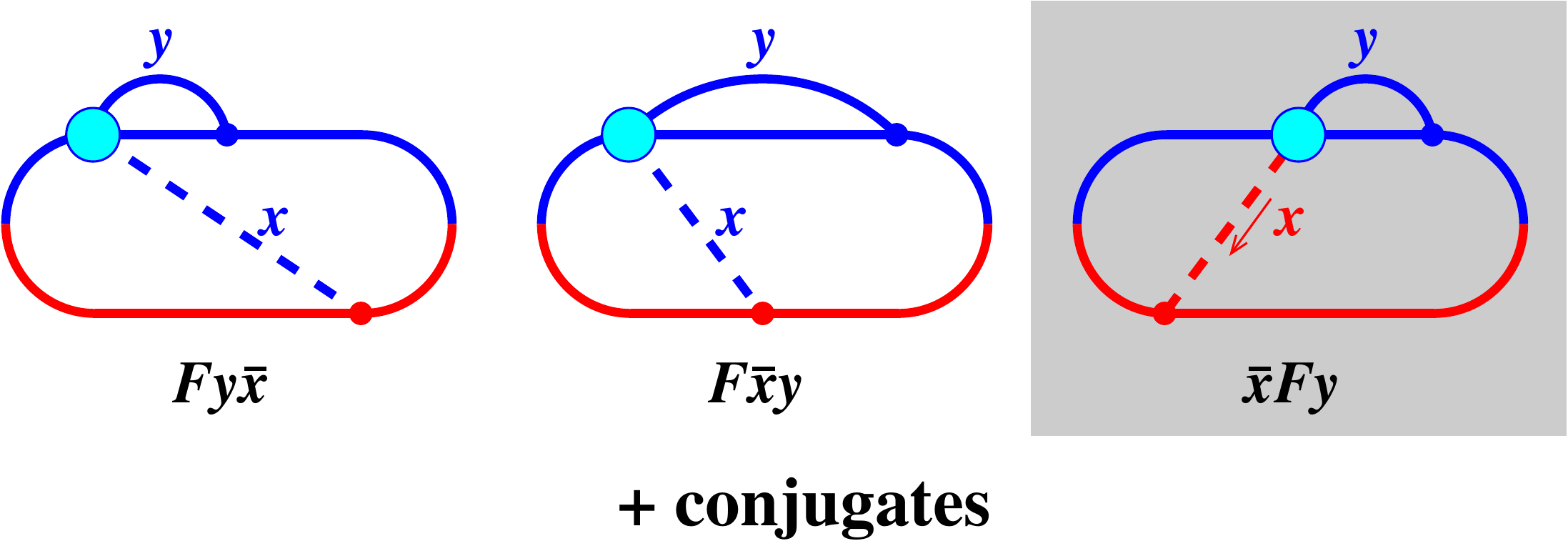}
  \caption{
     \label{fig:FdiagsVirtI}
     Like fig.\ \ref{fig:FdiagsReal} but for Class I virtual corrections
     to single splitting $g{\to}gg$.  Our terminology
     ``Class I'' \cite{qcd}
     means that (i) $y$ should be integrated
     over $0 < y < 1{-}x$ for these diagrams and
     (ii) $x \to 1{-}x$ generates another, distinct set of diagrams.
  }
\end {center}
\end {figure}

\begin {figure}[t]
\begin {center}
  \includegraphics[scale=0.5]{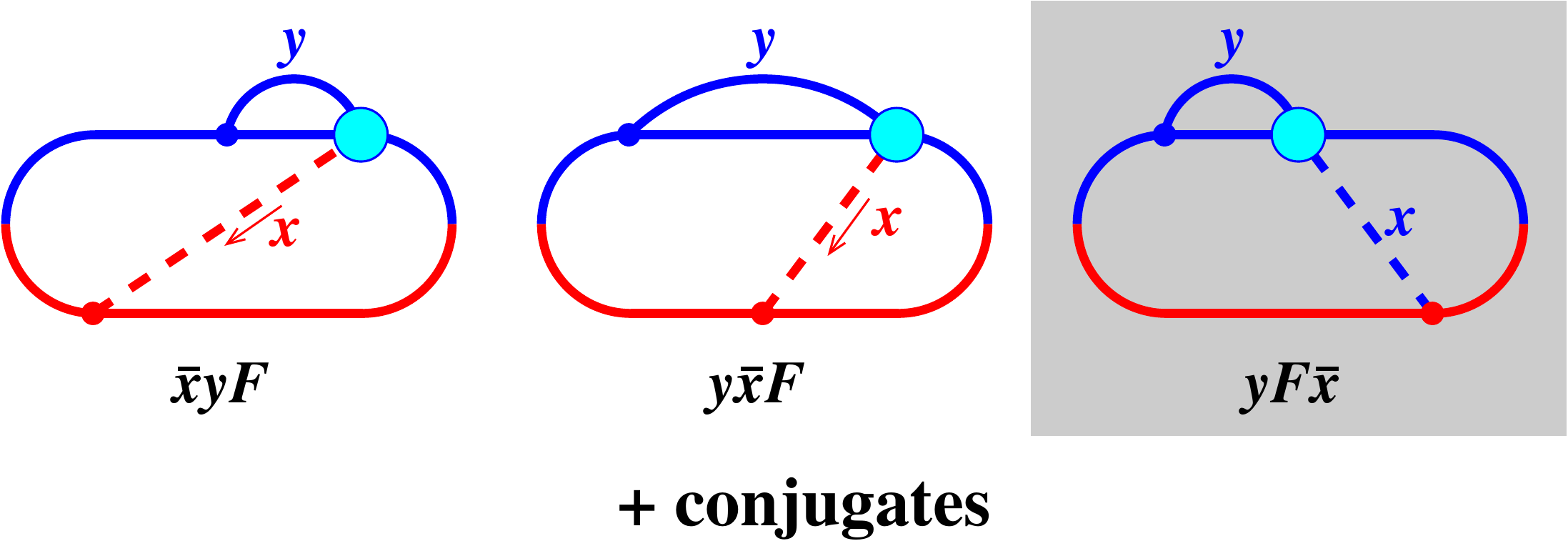}
  \caption{
     \label{fig:FdiagsVirtII}
     Like figs.\ \ref{fig:FdiagsReal} and \ref{fig:FdiagsVirtI}
     but for Class II virtual corrections
     to single splitting $g{\to}gg$.  Our terminology
     ``Class II'' \cite{qcd}
     means that (i) $y$ should be integrated
     over $0 < y < 1$ for these diagrams and
     (ii) the diagrams are symmetric under
     $x \to 1{-}x$.
  }
\end {center}
\end {figure}

\begin {figure}[t]
\begin {center}
  \includegraphics[scale=0.5]{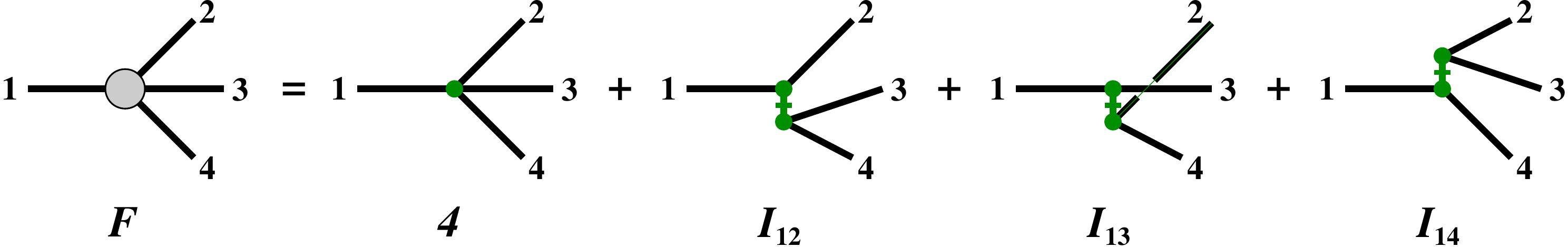}
  \caption{
     \label{fig:Fvertex}
     The meaning of the circular blob (``F'')
     in terms of the fundamental 4-gluon
     vertex (``4'') and LCPT instantaneous longitudinal gluon exchange in
     various channels (${\rm I}_{12}$, ${\rm I}_{13}$, ${\rm I}_{14}$).
     The green color here has no meaning other than to highlight the
     interactions that make up the circular blob.
  }
\end {center}
\end {figure}

The goal of this paper, then, is to analyze all remaining gluonic QCD diagrams.
These involve either (i) instantaneous longitudinal gluon exchange in
LCPT or (ii) fundamental 4-gluon vertices.
A complete list of such diagrams is depicted by
figs.\ \ref{fig:FdiagsReal}--\ref{fig:FdiagsVirtII}, plus additionally
diagrams obtained by replacing $x {\to} 1{-}x$ in fig.\ \ref{fig:FdiagsVirtI}.
Each circular blob in the diagrams
represents the sum of a fundamental 4-gluon vertex plus
all possible channels for a longitudinal gluon exchange, as depicted
in fig.\ \ref{fig:Fvertex}.

In naming time-ordered diagrams,
such as $xy\bar y\bar x$ in fig.\ \ref{fig:previous}a,
we follow refs.\ \cite{2brem,qcd} and
refer to the gluons in order of the time when they were emitted.
The absence or presence of a bar over a letter indicates whether the
emission at that time was in the amplitude or conjugate amplitude.
As in fig.\ \ref{fig:Fvertex}, we will use ``4'' to denote a
fundamental 4-gluon vertex and use ``I'' to denote an instantaneous
exchange of a longitudinal gluon in LCPT.  Effectively, these are both different
types of four-point interactions of transversely-polarized gluons.
When combined together, as
in the circular blobs of
figs.\ \ref{fig:FdiagsReal}\textbf{--}\ref{fig:Fvertex},
we will refer to sum with the letter ``F,'' which is intended to
evoke the word ``four.''

\begin {figure}[t]
\begin {center}
  \includegraphics[scale=0.5]{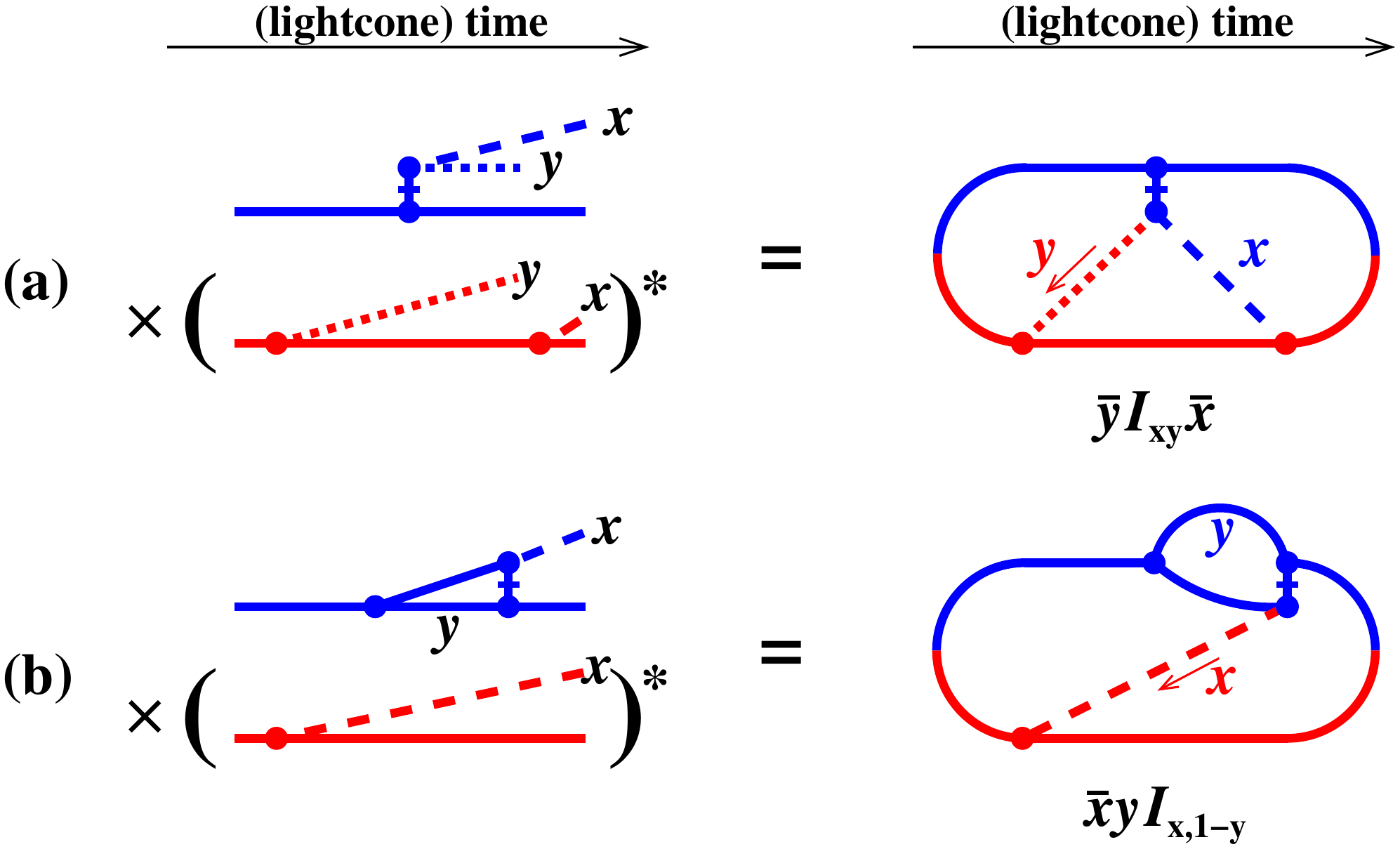}
  \caption{
     \label{fig:Itypes}
     Specific examples of instantaneous diagrams (depicted here in just
     one of the three possible channels)
     that are included
     in the (a) $\bar y F\bar x$ diagram of fig.\ \ref{fig:FdiagsReal}
     and (b) $\bar x y F$ diagram of fig.\ \ref{fig:FdiagsVirtII}.
  }
\end {center}
\end {figure}

It's worth noting that there are two different types of processes
where instantaneous longitudinal gluon exchange  plays a role
in figs.\ \ref{fig:FdiagsReal}--\ref{fig:FdiagsVirtII}.
One is by mediating $1{\to}3$ gluon pair creation processes as in
fig.\ \ref{fig:examples}.  All of the instantaneous vertices included
in figs.\ \ref{fig:FdiagsReal} and \ref{fig:FdiagsVirtI}
are of this type.  Because of the compact way the diagrams are drawn,
this may not be visually obvious in some cases, such as
the $\bar y F \bar x$ diagram of fig.\ \ref{fig:FdiagsReal},
but the interpretation can be clarified by redrawing the diagrams as a product
of an amplitude and conjugate amplitude, as in fig.\ \ref{fig:Itypes}a.
In contrast, the instantaneous vertices included in fig.\ \ref{fig:FdiagsVirtII}
represent $2{\to}2$ final-state rescattering corrections
(via longitudinal gluon exchange) to
the leading-order $g{\to}gg$ single-splitting process,
as depicted
for $\bar x y F$ in fig. \ref{fig:Itypes}b.

% ............................................................................

\subsection{Assumptions and Simplifications}

We will make the same simplifying
assumptions made for other diagrams in ref.\ \cite{qcd}
(and throughout the program of
refs.\ \cite{2brem,seq,dimreg,4point,QEDnf,qedNfstop,qcd}
for treating overlaps of successive hard splittings).
We work in the theorist's limit of an infinite, static,
homogeneous QCD medium.%
\footnote{
   More specifically, we assume that the QCD medium is approximately
   homogeneous over distances and times of order the formation length,
   which is parametrically of order $\sqrt{E/\hat q}$ for the case of
   quasi-democratic (i.e.\ not soft emission) splittings in an
   infinite QCD medium.
}
In this context, we take the high-energy
limit and make the corresponding high-energy approximation that
the relevant interactions with the medium can be described by
the medium parameter $\hat q$, defined as the proportionality
constant in the formula $\langle p_\perp^2 \rangle = \hat q L$ for
the typical $p_\perp$ picked up by a high-energy particle
traversing distance $L$ in the medium (for $L$ large compared
to the mean-free path for scattering from the medium).
We formally treat the running coupling $\alphas(\mu)$
as small at scales associated with the splitting vertices for
high-energy particles.%
\footnote{
  That scale is parametrically $\mu \sim (\hat q E)^{1/4}$ for
  quasi-democratic splittings in an infinite QCD medium.
}
Throughout, we will only consider rates that have been integrated
over the transverse momenta $p_\perp$ of the final-state daughters of
the $g{\to}ggg$ or $g{\to}gg$ splitting process.
We will also work in the large-$\Nc$ limit (where $\Nc$ is the number of
colors), which drastically simplifies
color dynamics for the overlap calculation.%
\footnote{
   A calculation of $1/\Nc^2$ corrections to previously-calculated
   $g{\to}ggg$ interference diagrams can be found in ref.\ \cite{1overN},
   which suggests that $\Nc \to \infty$ is a moderately good
   approximation.  (With caveats best left to ref.\ \cite{1overN} to
   describe, $1/\Nc^2$ corrections were $\le 17$\% for the
   processes studied there.)  A more general discussion of how the overlap
   calculation could be performed directly for $\Nc=3$
   may be found in ref.\ \cite{color}, though numerical implementation
   might be challenging.
}

% ............................................................................

\subsection{Outline}

Our strategy in this paper will be to first, in section \ref{sec:gTOggg},
evaluate the real double-splitting
$g{\to}ggg$ diagrams of fig.\ \ref{fig:FdiagsReal} by adapting the
calculations of ref.\ \cite{4point}, which were for those diagrams
that have fundamental 4-gluon vertices.
In section \ref{sec:virt}, we then transform
those $g{\to}ggg$ results to obtain results for the virtual diagrams of
figs.\ \ref{fig:FdiagsVirtI} and \ref{fig:FdiagsVirtII} by using the
diagrammatic techniques of ``front-end'' and ``back-end'' transformations
that were developed in ref.\ \cite{QEDnf} in the context of large-$\Nf$
QED and later applied to gluon splitting processes in ref.\ \cite{qcd}.
A detailed summary of
our final formulas for the effect of overlapping formation times
on splitting rates is given in appendix \ref{app:summary}, in a format
allowing easy integration with the earlier ``nearly-complete'' results
of ref.\ \cite{qcd}.
The goal of this paper is merely to obtain formulas for the relevant rates.
Our short conclusion in
section \ref{sec:conclusion} 
briefly references where one must
go from here to evaluate the relative importance of the new contributions.

% ============================================================================

\section{$g{\to}ggg$ processes with instantaneous interactions}
\label {sec:gTOggg}

\subsection{The $F\bar y\bar x$ diagram}

For a concrete start, we now discuss how to generalize earlier
results for the $4\bar y\bar x$ interference diagram of fig.\
\ref{fig:4point}a to include instantaneous diagrams and so obtain
the more general $F\bar y\bar x$ diagram of fig.\ \ref{fig:FdiagsReal}.

% ............................................................................

\subsubsection{Large-$\Nc$ color routings}

\begin {figure}[t]
\begin {center}
  \includegraphics[scale=0.5]{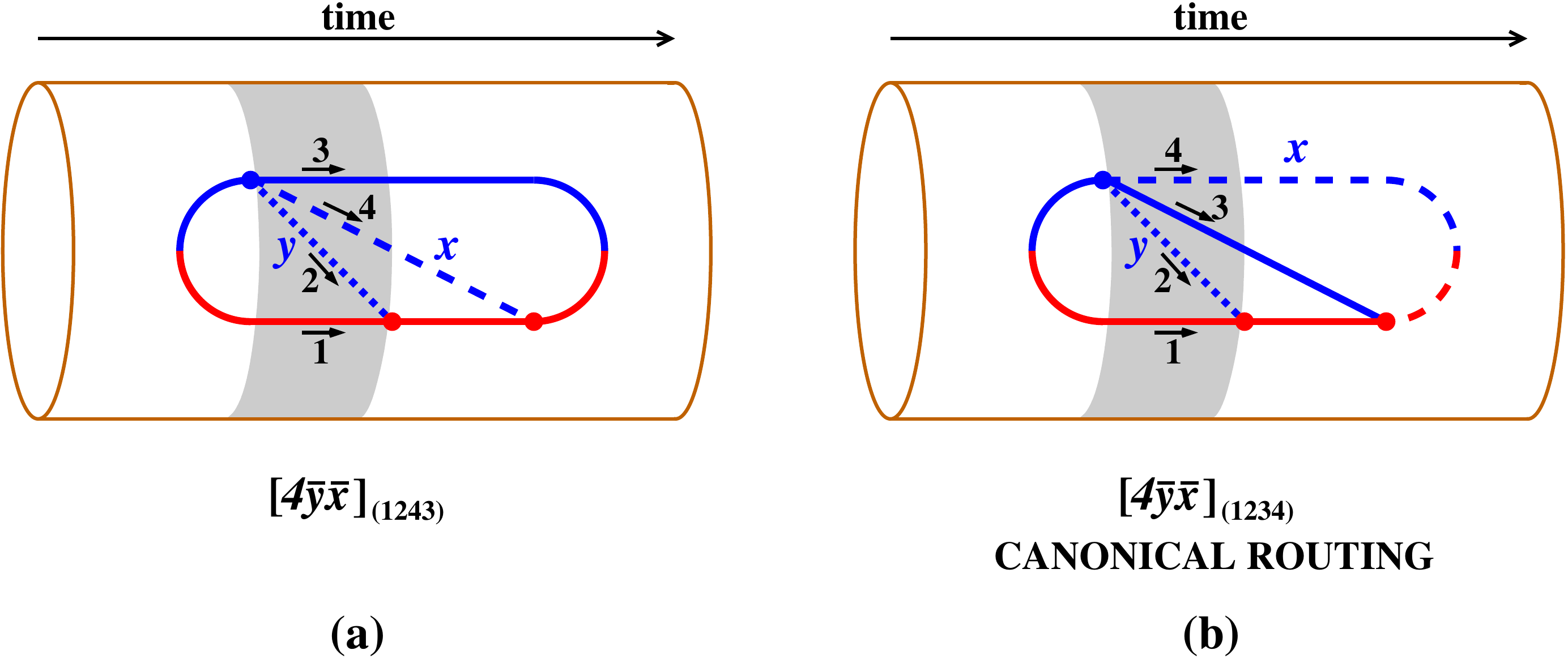}
  \caption{
     \label{fig:4yxRoute}
     The two large-$\Nc$ color routings of the $4\bar y\bar x$ diagram,
     drawn on a time-ordered cylinder.  These particular diagrams are
     drawn completely on the front side of the cylinder.
     The shaded region shows the times where four particles are present
     in the interference diagram (three in the amplitude plus one in
     the conjugate amplitude).  Numbering of the lines in that region
     is according to the convention (\ref{eq:x1234}).
  }
\end {center}
\end {figure}

One effect of taking the large-$\Nc$ limit to simplify color dynamics
is that certain types of interference diagrams get contributions
from more than one way to route large-$\Nc$ color in those diagram
\cite{seq,4point}.%
\footnote{
   See, in particular, section 2.2.1 of ref.\ \cite{seq} and
   sections 2.2 and 3.3 of ref.\ \cite{4point}.
}
A simple way to picture different large-$\Nc$ color routings for
a time-ordered diagram is (following refs.\ \cite{seq,4point}) to draw
the diagram without crossing lines on a cylinder, where time runs
along the length of the cylinder.  Fig.\ \ref{fig:4yxRoute}a,
adapted from ref.\ \cite{4point},%
\footnote{
  \label{foot:x1234warning}
  See figs.\ 11a and 12a of ref.\ \cite{4point}.  An important and
  potentially confusing difference is that, unlike ref.\ \cite{4point},
  our convention here is that the lines are always
  numbered according to (\ref{eq:x1234}).
}
gives
one example for the $4\bar y\bar x$ diagram.
There is a different large-$\Nc$ color routing for each different
way you can choose which high-energy particles neighbor each
other as one circles around the circumference of the cylinder.
There are exactly two different possibilities for the $4\bar y\bar x$ diagram,
both shown in fig.\ \ref{fig:4yxRoute}.
We must separately analyze these color routings because
the medium-averaged interactions of the high-energy particles
with the medium during 4-particle evolution
(the gray region) is different in the two case.
That's because, in the large-$\Nc$ limit, medium interactions
of gluon lines are correlated only between neighbors.

Following earlier work, we number the lines in these figures according
to the longitudinal momentum fractions of the lines as
\begin {equation}
  (x_1,x_2,x_3,x_4) = (-1,y,1{-}x{-}y,x) .
\label {eq:x1234}
\end {equation}
With this convention, the order of particles going around the cylinder
in the gray (4-particle evolution) section of
fig.\ \ref{fig:4yxRoute}b is (1234), which means that any pair of
particles are neighbors except for the pairs 1,3 and 2,4.%
\footnote{
  We need not consider the order of particles going around
  the circle in the 3-particle evolution parts of fig.\ \ref{fig:4yxRoute}
  because, for 3-particle evolution, all three particles are neighbors of
  each other.  This is related to the fact that, even for finite $\Nc$,
  there is no interesting color dynamics associated with 3-particle
  evolution in this application.
  See, for example, the arguments in section 2.3--2.4 of ref.\ \cite{2brem}
  or the discussion, in the context of the $\hat q$ approximation, of
  ref.\ \cite{Vqhat}.
}
In contrast, the particle order for fig.\ \ref{fig:4yxRoute}a is (1243).%
\footnote{
\label{foot:numbering}
   Our numbering convention here is different from fig.\ 11 of
   ref.\ \cite{4point}.  Here, we always number the lines according
   to the momentum fractions as in (\ref{eq:x1234}).  In contrast,
   fig.\ 11 of ref.\ \cite{4point} always numbers the lines
   so that they appear in the order (1234) and instead
   permutes which values $(x_1,x_2,x_3,x_4)$ refer to.
   In the end, it amounts to the same thing.
}
The
contributions of these two color routings of $4\bar y\bar x$ to the differential
rate $d\Gamma/dx\,dy$ are related to each other by simply interchanging
$x_3 \leftrightarrow x_4$, which is equivalent to $x \to 1{-}x{-}y$.
It's our custom to refer to the routing (1234) as our
``canonical'' routing in this context
and then obtain the result for the
other routing by substitution.  Henceforth, we'll refer to
the contribution to the rate from a canonical routing as
$[d\Gamma/dx\,dy]^{\rm canon}$.  So, for the $4\bar y\bar x$ diagram,%
\footnote{
  In ref.\ \cite{4point}, the two routings of fig.\ \ref{fig:4yxRoute}
  were called (a) $4\bar y\bar x_1$ and (b) $4\bar y\bar x_2$.
  The contribution from the canonical routing was then called
  $[d\Gamma/dx\,dy]_{4\bar y\bar x_2}$.  We write that as
  $[d\Gamma/dx\,dy]_{4\bar y\bar x}^{\rm canon}$ here because the new notation
  seems less obscure.
}
\begin {equation}
  \left[ \frac{d\Gamma}{dx\,dy} \right]_{4\bar y\bar x}
  =
  \left[ \frac{d\Gamma}{dx\,dy} \right]_{4\bar y\bar x}^{\rm canon}
  +
  ~(x \to 1{-}x{-}y) .
\end {equation}

\begin {figure}[t]
\begin {center}
  \includegraphics[scale=0.43]{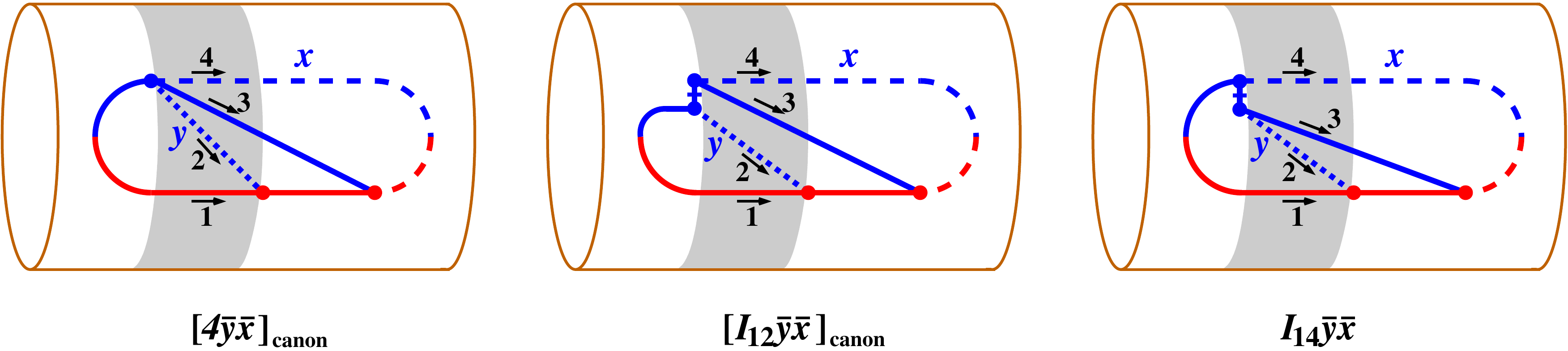}
  \caption{
     \label{fig:FyxCanon}
     Like fig.\ \ref{fig:4yxRoute}b but here including all instantaneous
     diagrams that contribute to the canonical routing (1234) and so
     to $[d\Gamma/dx\,dy]_{F\bar y\bar x}^{\rm canon}$.
  }
\end {center}
\end {figure}

Let's now do the same but also include instantaneous diagrams:
\begin {equation}
  \left[ \frac{d\Gamma}{dx\,dy} \right]_{F\bar y\bar x}
  =
  \left[ \frac{d\Gamma}{dx\,dy} \right]_{F\bar y\bar x}^{\rm canon}
  +
  ~(x \to 1{-}x{-}y) .
\end {equation}
The complete set of 4-point plus instantaneous color-routed
diagrams that can contribute to the canonical color routing (1234) is shown
on the cylinder in fig.\ \ref{fig:FyxCanon}.
In the diagram labels, we have not written a
``canon'' subscript on ``$I_{14}\bar y\bar x$'' because there
is only {\it one} possible large-$\Nc$ color routing of that
particular time-ordered diagram --- the canonical one.
There is no way to obtain the canonical large-$\Nc$
color routing using $I_{13}$.%
\footnote{
  You can't draw a {\it canonically-routed} (1234)
  $I_{13}\bar y\bar x$ diagram
  on the cylinder without crossing any lines.
  In the large-$\Nc$ limit, $I_{13}$ only contributes to the other
  color routing (1243) of $F\bar y\bar x$.
}

% ............................................................................

\subsubsection{Diagrammatic rule for longitudinal gluon exchange}

The $4\bar y\bar x$ diagram was previously calculated in ref.\ \cite{4point}.
To evaluate the other diagrams of fig.\ \ref{fig:FyxCanon}, we will
leverage the previous result by only computing in this paper the
{\it relative} overall factors of the three diagrams.  Then we will
adjust the overall factor of the earlier $4\bar y\bar x$ result
correspondingly.  The relative factors include the effects
of helicity contractions, color contractions, and longitudinal
momentum fraction ($x_i$) dependence associated with the different
four-gluon interactions $F$ in the different diagrams of
fig.\ \ref{fig:FyxCanon}.
Everything else about the
diagrams (the 3-gluon vertices, the evolution of the high-energy
particles in the medium) is the same.

\begin {figure}[t]
\begin {center}
  \begin{picture}(325,110)(0,0)
    \put(0,0){\includegraphics[scale=0.5]{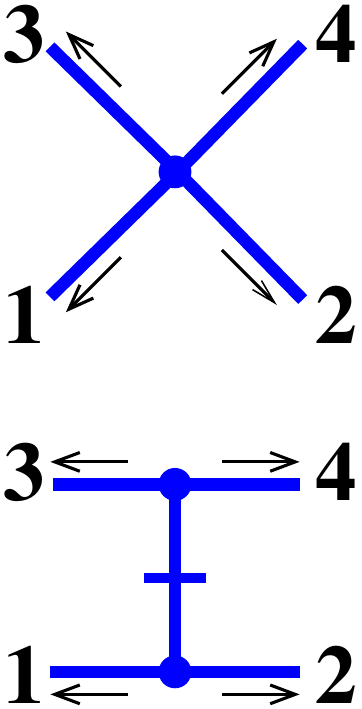}}
    % 4-fluon vertex rule
    \put(75,91){
      $\displaystyle{
        -i g^2 \Bigl\{
           f^{abe} f^{cde} ( \delta_{h_1,-h_3} \delta_{h_2,-h_4}
                         - \delta_{h_1,-h_4} \delta_{h_2,-h_3} )
      }$
    }
    \put(103,74){
      $\displaystyle{
           +
           f^{ace} f^{dbe} ( \delta_{h_1,-h_4} \delta_{h_2,-h_3}
                         - \delta_{h_1,-h_2} \delta_{h_3,-h_4} )
      }$
    }
    \put(103,57){
      $\displaystyle{
           +
           f^{ade} f^{bce} ( \delta_{h_1,-h_2} \delta_{h_3,-h_4}
                         - \delta_{h_1,-h_3} \delta_{h_2,-h_4} )
         \Bigr\}
      }$
    }
    % longitudinal gluon exchange rule
    \put(75,18){
      $\displaystyle{
           i g^2
           f^{abe} f^{cde} \delta_{h_1,-h_2} \delta_{h_3,-h_4} \,
           \frac{(p_1^+-p_2^+)(p_3^+-p_4^+)}{(p_1^++p_2^+)^2}
      }$
    }
  %\put(0,0){.}
  %\put(0,110){.}
  %\put(325,0){.}
  %\put(325,110){.}
  \end{picture}
  \caption{
     \label {fig:rules}
     Ordinary Feynman rule results in lightcone gauge
     for the four-gluon interactions appearing in our ``$F$'' diagrams.
     By ordinary Feynman rules, we mean that (i) there is no direction of time
     in the above diagrams, (ii) we are not yet taking the
     large-$\Nc$ limit nor yet separating out different large-$\Nc$ color
     routings, and (iii) we are only talking about the interaction in an
     amplitude and are not thinking here about the conjugate amplitude.
     As in the rest of the text, unbarred lines refer to transversely
     polarized gluons, and the barred line is a longitudinally polarized
     gluon.  $(a,b,c,d)$ are the adjoint-index colors of gluons
     $(1,2,3,4)$ respectively, and $(h_1,h_2,h_3,h_4)$ are the corresponding
     helicities $\pm$ flowing in the direction of the arrows.
     The Kronecker deltas arise from dot products
     $\beps_i\cdot\beps_j = \delta_{h_i,-h_j}$ of the
     two-dimensional unit polarization vectors of (\ref{eq:epshelicity}).
  }
\end {center}
\end {figure}

The easiest way to compare the different four-gluon interactions is
to forget about time-ordered perturbation theory for a moment and
just think about Feynman rules.
These are shown for light-cone gauge $A^+{=}0$ in fig.\ \ref{fig:rules},
where we follow our convention that unbarred lines represent
transversely polarized gluons and the barred line represents a
longitudinally polarized gluon.  One may take the rule for
longitudinal gluon exchange from the literature.%
\footnote{
  See, for example, fig.\ 54 of ref.\ \cite{BL}.
  This is the same as our
  longitudinal gluon exchange rule in fig.\ \ref{fig:rules}
  after a few adjustments.  (i) The labeling of the particles is
  different.  (ii) Presumably a typographic error: their denominator
  $(p_c^+ + p_b^+)$ should be $(p_c^++p_d^+)^2$.  (iii) Though they draw arrows
  on their gluon lines indicating the same convention for gluon momentum
  flow as our fig.\ \ref{fig:rules}, they, unlike us, do not adopt
  this same convention for helicity flow.  So their $\eps_a^* \cdot \eps_b$
  and $\eps_d^*\cdot \eps_b$ correspond to what we would call (if we
  used their labeling of lines but our helicity flow convention)
  $\eps_a^* \cdot \eps_b^* = -\delta_{h_a,-h_b}$ and
  $\eps_d^* \cdot \eps_b^* = -\delta_{h_d,-h_b}$.
  (iv) Ordinary Feynman rules correspond to a perturbative expansion of
  $e^{iS}$, where $S$ is the action.  Our fig.\ \ref{fig:rules}
  corresponds to contributions to $i S_{\rm eff}$, where $S_{\rm eff}$
  is the effective action after one integrates out longitudinal
  polarizations.  In contrast, the rules of ref.\ \cite{BL} are
  for the Hamiltonian.  For these interactions, there is a relative
  minus sign between $S_{\rm eff}$ and $H_{\rm eff}$, and so our
  rules are $-i$ times their rules.
  One may similarly compare our fig.\ \ref{fig:rules}
  to tables 2 and  3 of ref.\ \cite{BPP},
  where the overall sign and momentum dependence are the same as
  ref.\ \cite{BL},
  but the overall normalization is more difficult to compare because
  ref.\ \cite{BPP} uses unusual normalization conventions.
}
But, since some
of the LCPT literature has confusing normalization or sign issues,
we will take a moment here to briefly review the derivation.

In lightcone gauge, the
basis $\eps_{(\lambda)}$ for transverse polarizations of a gauge boson
with
4-momentum $p$ is given by
\begin {equation}
   (\eps^+,\eps^-,\beps)_{(\lambda)}
   =
   \Bigl( 0, \frac{\beps_{(\lambda)}\cdot\p}{p^+},
          \beps_{(\lambda)} \Bigr) ,
\label {eq:epsT}
\end {equation}
where
$\beps_{(\lambda)}$ is any basis of unit spatial vectors
for the $xy$-plane.
For a helicity basis, one may choose, for example,
\begin {equation}
   \beps_{(\pm)} = \bigl( \tfrac{1}{\sqrt2}, \pm \tfrac{i}{\sqrt2} \bigr).
\label {eq:epshelicity}
\end {equation}
Here and throughout, boldface letters like
$\beps$ and $\p$ will denote the projection of vectors onto the
$xy$-plane.
Our convention for lightcone coordinates
is that
$v^\pm \equiv (v^0 \pm v^3)/\sqrt2$. So the 4-vector dot product
[in $({+}{-}{-}{-})$ metric convention] is
$u\cdot v = u^+ v^- + u^- v^+ - {\bm u}\cdot{\bm v}$,
and $v_\pm = v^\mp$.
The longitudinal polarization is
\begin {equation}
   (\eps^+,\eps^-,\beps)_{\rm L}
   =
   ( 0, 1, {\bm 0} ) .
\end {equation}
All three 4-vector basis polarizations are orthogonal to each other.
The transverse polarizations are furthermore orthogonal to
4-vector $p$ and normalized so
that $(\eps_{(\lambda)}^*)^\mu(\eps_{(\lambda')})_\mu = -\delta_{\lambda\lambda'}$.

The lightcone gauge propagator is (ignoring $i\varepsilon$ prescriptions
for now) 
\begin {equation}
     G^{\mu\nu}(q) =
     -\frac{i}{q^2} \Bigl[
       g^{\mu\nu} - \frac{q^\mu n^\nu + q^\nu n^\mu}{q\cdot n}
     \Bigr] ,
\label {eq:Glightcone}
\end{equation}
where it's convenient to rewrite lightcone gauge $A^+=0$ as
$n\cdot A=0$ with $(n^+,n^-,{\bm n}) = (0,1,{\bm 0})$.
The propagator (\ref{eq:Glightcone}) may be recast into the form
\begin {subequations}
\label {eq:GLTdecomp}
\begin {equation}
     G^{\mu\nu}(q) = G^{\mu\nu}_{\rm T}(q) + G^{\mu\nu}_{\rm L}(q)
\end{equation}
with
\begin {equation}
    G^{\mu\nu}_{\rm T}(q) =
    \frac{i}{q^2}
    \sum_\lambda \eps_{(\lambda)}^\mu(q) \, \eps_{(\lambda)}^{\nu*}(q) ,
    \qquad
    G^{\mu\nu}_{\rm L}(q) = \frac{i}{(q\cdot n)^2} \,  n^\mu n^\nu . 
\end {equation}
\end {subequations}
Note that $q\cdot n = q^+$.

The rule for the longitudinally polarized
gluon exchange in fig.\ \ref{fig:rules}
comes from applying normal Feynman rules
but including only
the longitudinal piece $G_{\rm L}$ of
the lightcone propagator for the exchanged gluon.
The result that this rule is independent of
any $p^-$ is the reason why (after Fourier transformation to
coordinate space) the
interaction is instantaneous in lightcone time $x^+$.
It is also local in ${\bm x} \equiv (x^1,x^2)$ but is non-local in $x^-$.

% ............................................................................

\subsubsection{The color routings and contractions for fig.\ \ref{fig:FyxCanon}}

Now turn to the large-$\Nc$, canonically routed diagrams of
fig.\ \ref{fig:FyxCanon}.  In our convention for defining the
flow of momenta there, all of the arrows flow away from the
four-gluon interaction in the amplitude, matching the flow
convention of fig.\ \ref{fig:rules}.
Note that the fundamental 4-point vertex in fig.\ \ref{fig:rules}
can be written as
\begin {multline}
        -i g^2 \Bigl\{
           \tfrac12
           f^{abe} f^{cde} ( \delta_{h_1,-h_3} \delta_{h_2,-h_4}
                         - \delta_{h_1,-h_4} \delta_{h_2,-h_3} )
           +
           f^{ade} f^{bce} ( \delta_{h_1,-h_2} \delta_{h_3,-h_4}
                         - \delta_{h_1,-h_3} \delta_{h_2,-h_4} )
         \Bigr\}
\\
   + (\mbox{interchange particles 3 and 4}) ,
\label {eq:4vertex}
\end {multline}
and remember that the two different color routings of the $4\bar y\bar x$
diagram are related by interchange of particles 3 and 4.
In ref.\ \cite{4point},
the piece of our (\ref{eq:4vertex}) that contributes to the
canonical large-$\Nc$ color routing of the $4\bar y\bar x$ diagram
in fig.\ \ref{fig:FyxCanon}
was found to be the first term in (\ref{eq:4vertex}):
\begin {multline}
  [4\bar y\bar x]_{\rm canon}
  \propto
      -i g^2 \Bigl\{
           \tfrac12
           f^{abe} f^{cde} ( \delta_{h_1,-h_3} \delta_{h_2,-h_4}
                         - \delta_{h_1,-h_4} \delta_{h_2,-h_3} )
\\
           +
           f^{ade} f^{bce} ( \delta_{h_1,-h_2} \delta_{h_3,-h_4}
                         - \delta_{h_1,-h_3} \delta_{h_2,-h_4} )
       \Bigr\} .
\label {eq:4prop1}
\end {multline}

If one ignored color routing, the $I_{12}$ interaction of
$I_{12}\bar y\bar x$ would give
\begin {equation}
           i g^2
           f^{abe} f^{cde} \delta_{h_1,-h_2} \delta_{h_3,-h_4} \,
           \frac{(p_1^+-p_2^+)(p_3^+-p_4^+)}{(p_1^++p_2^+)^2} .
\label {eq:I12}
\end {equation}
This single term is symmetric under exchange of particles 3 and 4,
and we find that each large-$\Nc$ color routing corresponds to
half of it:
\begin {subequations}
\label {eq:Iprop1}
\begin {equation}
  [I_{12}\bar y\bar x]_{\rm canon}
  \propto
           \tfrac12 \,
           i g^2
           f^{abe} f^{cde} \delta_{h_1,-h_2} \delta_{h_3,-h_4} \,
           \frac{(p_1^+-p_2^+)(p_3^+-p_4^+)}{(p_1^++p_2^+)^2} .
\end {equation}
Finally, there is no color routing issue for the $I_{14}\bar y\bar x$
diagram, so we can convert the full (\ref{eq:I12}) for $I_{12}$
to $I_{14}$ by switching the labels of particles 2 and 4:
\begin {equation}
  I_{14}\bar y\bar x
  \propto
           i g^2
           f^{ade} f^{cbe} \delta_{h_1,-h_4} \delta_{h_3,-h_2} \,
           \frac{(p_1^+-p_4^+)(p_3^+-p_2^+)}{(p_1^++p_4^+)^2} .
\end {equation}
\end {subequations}

Eqs.\ (\ref{eq:4prop1}) and (\ref{eq:Iprop1}) are the only differences
in the evaluation of the three diagrams of fig.\ \ref{fig:FyxCanon}.
We'll find it convenient later on in this paper to have introduced
some short-hand notation for the various factors in these equations:
\begin {subequations}
\label {eq:shorthand}
\begin {equation}
   c_{12} \equiv f^{abe} f^{cde} ,
   \quad
   c_{13} \equiv f^{ace} f^{dbe} ,
   \quad
   c_{14} \equiv f^{ade} f^{bce} ;
\label {eq:cdef}
\end {equation}
\begin {equation}
   h_{12} \equiv \delta_{h_1,-h_2} \delta_{h_3,-h_4} ,
   \quad
   h_{13} \equiv \delta_{h_1,-h_3} \delta_{h_2,-h_4} ,
   \quad
   h_{14} \equiv \delta_{h_1,-h_4} \delta_{h_2,-h_3} ;
\label {eq:h1n}
\end {equation}
\begin {equation}
   i_{12} \equiv \frac{(x_1{-}x_2)(x_3{-}x_4)}{(x_1{+}x_2)^2} \,,
   \quad
   i_{13} \equiv \frac{(x_1{-}x_3)(x_4{-}x_2)}{(x_1{+}x_3)^2} \,,
   \quad
   i_{14} \equiv \frac{(x_1{-}x_4)(x_2{-}x_3)}{(x_1{+}x_4)^2} \,,
\label {eq:idef}
\end {equation}
\end {subequations}
where the $x_n$ are the $p^+$ momentum fractions defined by
$p_n^+ \equiv x_n P^+$, where $P$ is the 4-momentum of the initial particle
in the double-splitting process.%
\footnote{
  Given that the high-energy splitting processes are highly collinear in our
  application, one can just
  as well say that the $x_n$ are the energy fractions defined by
  $p_n^0 \equiv x_n E$, as we sometimes do elsewhere.
  But, in the context of LCPT, it's more precise and more general
  to say that they are $p^+$ momentum fractions.
}
With this notation, the relative factors that differ between
the three diagrams are
\begin {subequations}
\label {eq:diagnotation}
\begin {align}
  [4\bar y\bar x]_{\rm canon}
  &\propto
  - \tfrac12 c_{12} (h_{13}-h_{14}) - c_{14} (h_{12}-h_{13}),
\\
  [I_{12}\bar y\bar x]_{\rm canon}
  &\propto
  \tfrac12 c_{12} h_{12} i_{12} ,
\\
  I_{14}\bar y\bar x
  &\propto
           c_{14} h_{14} i_{14} ,
\end {align}
\end {subequations}
where we've now absorbed the common factor of $i g^2$ into the
joint proportionality.

For future reference, note that the $c_{1n}$ and $i_{1n}$ have been
defined in such a way that $(c_{12},c_{13},c_{14})$ and
$(i_{12},i_{13},i_{14})$ cyclically permute when the particle
labels $(234)$ are cyclically permuted. However, the definitions
pick up an additional minus sign when swapping just one pair
of particle labels.  For example, swapping particles 2 and 4 takes
$(c_{12},c_{13},c_{14}) \rightarrow (-c_{14},-c_{13},-c_{12})$ and
$(i_{12},i_{13},i_{14}) \rightarrow (-i_{14},-i_{13},-i_{12})$
and so takes
$(c_{12} i_{12}, c_{13} i_{13}, c_{14} i_{14}) \rightarrow
 (c_{14} i_{14}, c_{13} i_{13}, c_{12} i_{12})$.

In the calculation of rates, we will sum/average over final/initial
state helicities and colors, as we did for $[4\bar y\bar x]_{\rm canon}$
alone in ref.\ \cite{4point}.  To compare the relative rates among
our diagrams here, we now need to be explicit about what common factors
hidden in the common proportionality symbols depend on colors and
helicities.

Let's start by first focusing on color.
The color factors from the two 3-gluon vertices
in the $F\bar y\bar x$ diagrams of fig.\ \ref{fig:FyxCanon} are
proportional to $f^{abf} f^{cdf} = c_{12}$.  (Proportionality is enough
here.  Since they are the same for all three diagrams,
we do not have to keep track of the appropriate order
of the indices in the 3-gluon vertex $f$'s because that only affects the
{\it common} overall sign of those diagrams.)
Letting angle brackets
$\langle \cdots \rangle$ represent
summing/averaging over colors in this particular context,
one finds
\begin {equation}
   \langle c_{12} c_{12} \rangle = \CA^2 ,
   \qquad
   \langle c_{12} c_{13} \rangle = \langle c_{12} c_{14} \rangle
     = -\tfrac12 \CA^2
   .
\label {eq:ccOriginal}
\end {equation}
We then have%
\begin {subequations}
\label {eq:Iprop2}
\begin {align}
  [4\bar y\bar x]_{\rm canon}
  &\propto
  - \tfrac12 \langle c_{12} c_{12} \rangle (h_{13}-h_{14})
    - \langle c_{12} c_{14} \rangle (h_{12}-h_{13})
  \propto
  h_{12} - 2 h_{13} + h_{14} ,
\\
  [I_{12}\bar y\bar x]_{\rm canon}
  &\propto
  \tfrac12 \langle c_{12} c_{12} \rangle h_{12} i_{12}
  \propto
  h_{12} i_{12} ,
\\[5pt]
  I_{14}\bar y\bar x
  &\propto
  \langle c_{12} c_{14} \rangle h_{14} i_{14}
  \propto
  - h_{14} i_{14} ,
\end {align}
\end {subequations}
where we've absorbed a common factor of $\tfrac12 \CA^2$
into the second proportionality symbol of each line.

% ............................................................................

\subsubsection{Helicity contractions}

We now need to include the helicity dependence of the two 3-gluon
vertices and then sum/average over helicity.  In the notation of
refs.\ \cite{2brem,4point},
the 3-gluon vertices give
factors of%
\footnote{
  See, in particular, eq.\ (2.14) of ref.\ \cite{4point} for the
  $4\bar y\bar x$ diagram, or
  the earlier discussion of
  eq.\ (4.37) of ref.\ \cite{2brem} for the $xy\bar y\bar x$ diagram.
}
\begin {equation}
   \Bigl[
   \sum_{\bar h}
   {\cal P}^{\bar m}_{h_\ix \to \bar h, h_\yx}\!\bigl(1 \to 1{-}y,y\bigr) \,
   {\cal P}^{\bar n}_{\bar h \to h_\zx,h_\xx}\!\bigl(1{-}y \to 1{-}x{-}y,x\bigr) \,
   \Bigr]^* ,
\label {eq:PP}
\end {equation}
where the ${\cal P}$ are given in terms of
square roots of helicity-dependence vacuum
Dokshitzer-Gribov-Lipatov-Altarelli-Parisi (DGLAP) splitting functions.
The exact definitions can be found in ref.\ \cite{2brem},%
\footnote{
  See eqs.\ (4.32) and (4.35) of ref.\ \cite{2brem}.
}
where $\bm{\mathcal P}$ is defined as a 2-dimensional vector 
proportional to $(1,+i)$ or $(1,-i)$ depending on the specific
helicity transition.  The indices $\bar m$ and $\bar n$ in (\ref{eq:PP})
index the components of that vector.
$h_\ix$ is the helicity of the initial particle in the $g{\to}ggg$
splitting process; $(h_\xx,h_\yx,h_\zx)$ are the helicities of the
three daughters; and
$\bar h$ is the helicity of the unlabeled red line connecting the
two 3-gluon vertices in each of
the three diagrams of fig.\ \ref{fig:FyxCanon}.
With the numbering and flow direction conventions of the lines in
fig.\ \ref{fig:FyxCanon}, (\ref{eq:PP}) is
\begin {equation}
   \Sigma^{\bar m\bar n} \equiv
   \Bigl[
   \sum_{\bar h}
   {\cal P}^{\bar m}_{-h_1 \to \bar h, h_2}\!\bigl(1 \to 1{-}y,y\bigr) \,
   {\cal P}^{\bar n}_{\bar h \to h_3,h_4}\!\bigl(1{-}y \to 1{-}x{-}y,x\bigr) \,
   \Bigr]^* .
\end {equation}

Now let's sum/average over the daughter and parent helicities.
We'll denote that helicity sum/average using angle brackets as well.
In ref.\ \cite{4point}, the relevant average for the $4\bar y\bar x$
diagram was found to be%
\footnote{
  See the discussion of eqs.\ (2.14--2.16) of ref.\ \cite{4point}.
  We refer here to the $\zeta(x,y)$ of that reference as $\zeta_{(4)}$
  to distinguish it from the other $\zeta$'s we construct.
  The $\delta^{\bar m\bar n}$ dependence of our (\ref{eq:hsum4}) is just
  a consequence of transverse-plane rotational invariance after doing
  the helicity sums.
  The factor of $|x_1 x_2 x_3 x_4|^{-1/2}$ in our (\ref{eq:hsum4}) is
  merely a
  convenient normalization convention that was
  used for the definition of $\zeta$ in
  ref.\ \cite{4point}.
}
(using our notation here)
\begin {equation}
   \bigl\langle \Sigma^{\bar m\bar n} \, (h_{12} - 2 h_{13} + h_{14}) \bigr\rangle
       \, |x_1 x_2 x_3 x_4|^{-1/2}
   = \zeta_{(4)}(x,y) \, \delta^{\bar m\bar n}
\label {eq:hsum4}
\end {equation}
with
\begin {equation}
  \zeta_{(4)}(x,y) =
  \frac{
        2 x^2 - z^2 - (1{-}y)^4 + 2 y^2 z^2 - x^2 y^2
     }{(x y z)^2 (1{-}y)^3}
   \,.
\label {eq:zeta4}
\end {equation}
By repeating that calculation, we find now that the separate pieces of
(\ref{eq:hsum4}) are given by
\begin {equation}
   \langle \Sigma^{\bar m\bar n} \, h_{1k} \rangle \, |x_1 x_2 x_3 x_4|^{-1/2}
   = \zeta_{1k}(x,y) \, \delta^{\bar m\bar n}
\end {equation}
with%
\footnote{
  If desired, one may rewrite (\ref{eq:zetaij}) in terms of the
  $(x_1,x_2,x_3,x_4)$ variables of (\ref{eq:x1234}) as
  \[
    \zeta_{12} = \frac{ (x_1^2{+}x_2^2)(x_3^2{+}x_4^2) }
                     { (x_1 x_2 x_3 x_4)^2 |x_1{+}x_2|^3 } ,
    \quad
    \zeta_{13} = \frac{ (x_1{+}x_2)^4 + (x_1 x_3)^2 + (x_2 x_4)^2 }
                     { (x_1 x_2 x_3 x_4)^2 |x_1{+}x_2|^3 } ,
    \quad
    \zeta_{14} = \frac{ (x_1{+}x_2)^4 + (x_1 x_4)^2 + (x_2 x_3)^2 }
                     { (x_1 x_2 x_3 x_4)^2 |x_1{+}x_2|^3 } .
  \]
}
\begin {subequations}
\label {eq:zetaij}
\begin {align}
   \zeta_{12} &=
   \frac{(x^2{+}z^2)(1{+}y^2)}
        {(x y z)^2(1{-}y)^3} \,,
\\
   \zeta_{13} &=
   \frac{(1{-}y)^4 + z^2 + x^2 y^2}
        {(x y z)^2(1{-}y)^3} \,,
\\
   \zeta_{14} &=
   \frac{(1{-}y)^4 + x^2 + z^2 y^2}
        {(x y z)^2(1{-}y)^3} \,,
\end {align}
\end {subequations}
in terms of which
\begin {equation}
   \zeta_{(4)} = \zeta_{12} - 2\zeta_{13} + \zeta_{14} .
\label {eq:zeta4alt}
\end {equation}
So, after helicity summing/averaging, (\ref{eq:Iprop2}) becomes
\begin {subequations}
\label {eq:Iprop3}
\begin {align}
  [4\bar y\bar x]_{\rm canon}
  &\propto
  \bigl\langle \Sigma^{\bar m\bar n} \, (h_{12} - 2 h_{13} + h_{14}) \bigr\rangle
  \propto
  \zeta_{(4)} \,\delta^{\bar m\bar n} ,
\\
  [I_{12}\bar y\bar x]_{\rm canon}
  &\propto
  \langle \Sigma^{\bar m\bar n} \, h_{12} \rangle \, i_{12}
  \propto
  \zeta_{12} \, i_{12} \,\delta^{\bar m\bar n} ,
\\[5pt]
  I_{14}\bar y\bar x
  &\propto
  - \langle \Sigma^{\bar m\bar n} \, h_{14} \rangle \, i_{14}
  \propto
  - \zeta_{14} \, i_{14} \,\delta^{\bar m\bar n} .
\end {align}
\end {subequations}
From this, we see that the result for $4\bar y\bar x$
in ref.\ \cite{4point} can be converted
to a result for $F\bar y\bar x$ (which includes
instantaneous diagrams) by
\begin {equation}
  \left[ \frac{d\Gamma}{dx\,dy} \right]_{F\bar y\bar x}^{\rm canon}
  =
  \left\{
    \left[ \frac{d\Gamma}{dx\,dy} \right]_{4\bar y\bar x}^{\rm canon}
    ~\mbox{with}~
    \zeta_{(4)} \longrightarrow \zeta_{\rm (F)}
  \right\} ,
\label {eq:FyxSub}
\end {equation}
where
\begin {equation}
  \zeta_{\rm(F)} = \zeta_{(4)} + \zeta_{12} \, i_{12} - \zeta_{14} \, i_{14}
  =
   \zeta_{(4)}
   - \frac{(1{+}y)(z{-}x)}{(1{-}y)^2} \, \zeta_{12}
   - \frac{(1{+}x)(z{-}y)}{(1{-}x)^2} \, \zeta_{14} .
\end {equation}
We summarize the final formulas for this and all other rates
involving 4-gluon interactions in appendix \ref{app:summary}.

% ............................................................................

\subsubsection{The $\bar y\bar xF$ and $\bar y F \bar x$ diagrams}
\label {sec:yFx}

The color and helicity sums for the $\bar y \bar x F$ diagram are the
same as those for the $F\bar y\bar x$ diagram, and so the same
substitution $\zeta_{(4)} \to \zeta_{(F)}$ as in (\ref{eq:FyxSub})
can be made on the result for the canonical routing (1234) of
$\bar y\bar x 4$ from ref.\ \cite{4point}.

The $\bar y F \bar x$ diagram vanishes for the same reason as the
$\bar y 4 \bar x$ diagram in ref.\ \cite{4point}, which can be
argued from parity invariance of either the initial or final
3-particle evolution in this diagram.
(See section 3.2 of ref.\ \cite{4point}.)

% ----------------------------------------------------------------------------

\subsection {The $F\bar F$ diagram}

\begin {figure}[t]
\begin {center}
  \includegraphics[scale=0.43]{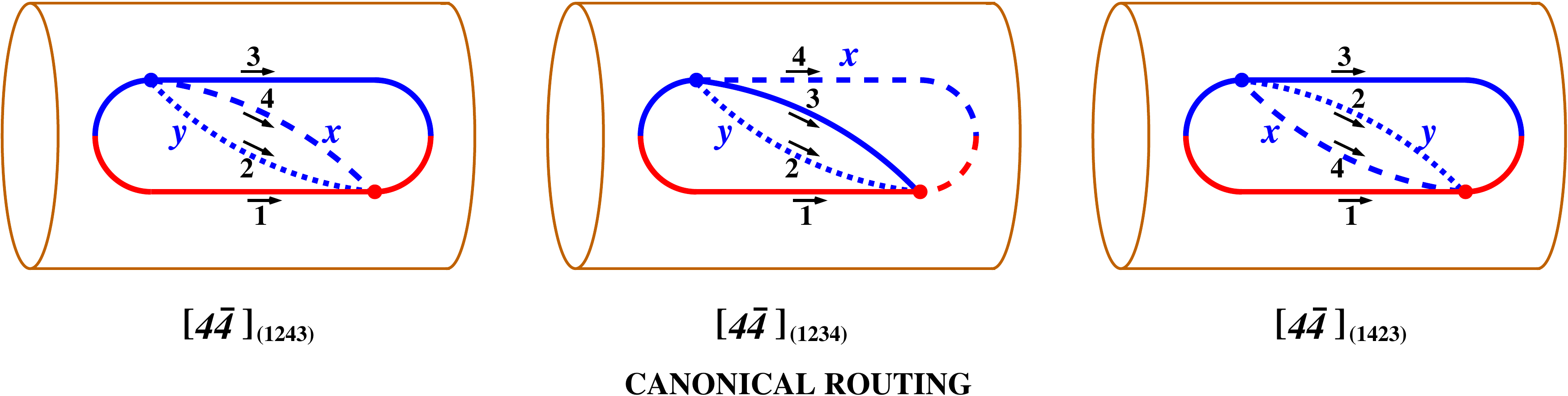}
  \caption{
     \label{fig:44route}
     The three large-$\Nc$ color routings of the $4\bar 4$ diagram,
     drawn with the same conventions as fig.\ \ref{fig:4yxRoute}
     except that here we have not bothered to shade the region of
     4-particle evolution.
     (This figure is adapted from fig.\ 14 of ref.\ \cite{4point}.)
  }
\end {center}
\end {figure}

\begin {figure}[t]
\begin {center}
  \includegraphics[scale=0.45]{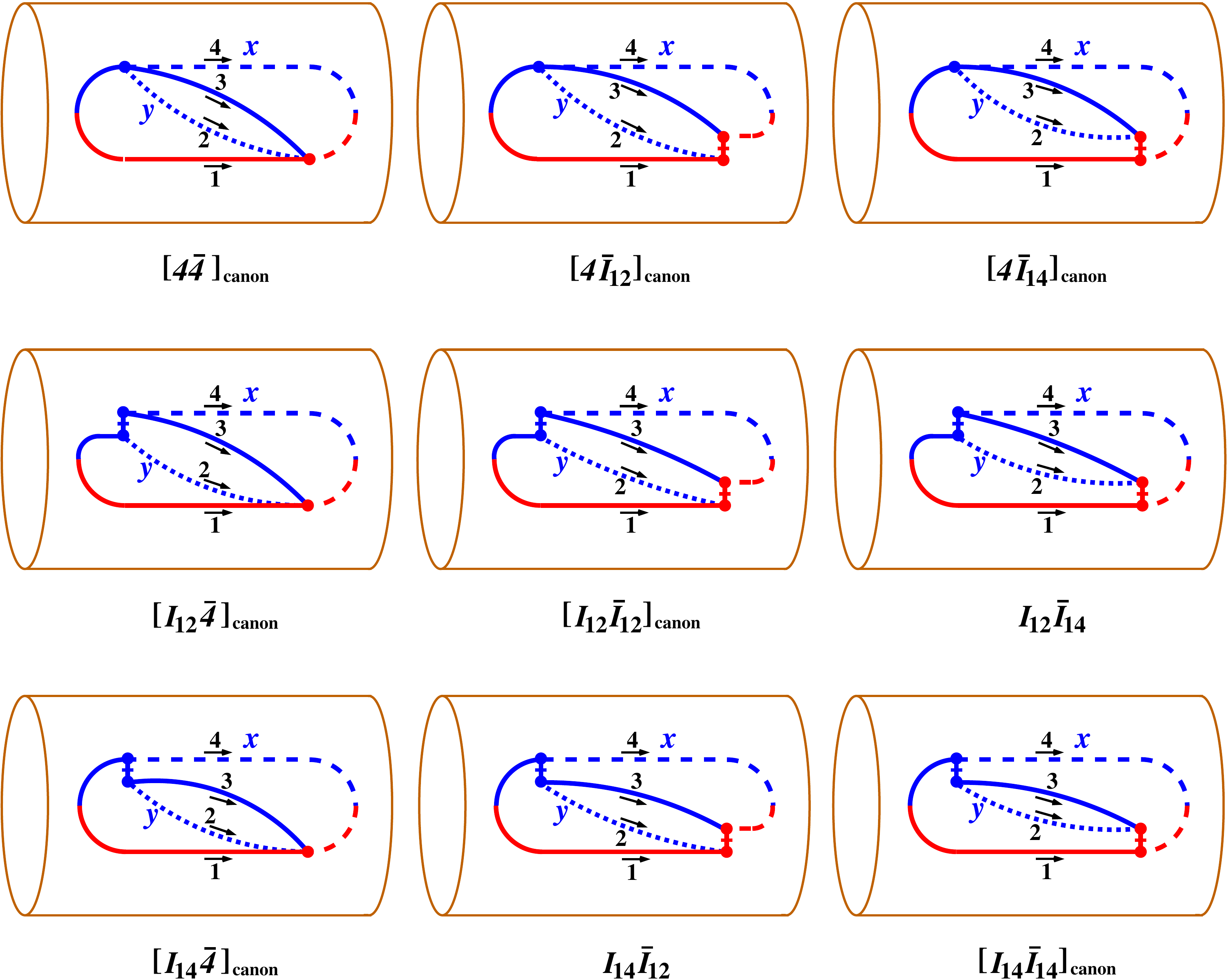}
  \caption{
     \label{fig:FFcanon}
     Like the middle diagram of
     fig.\ \ref{fig:44route} but here also including all instantaneous
     diagrams that contribute to the canonical routing (1234) and so
     to $[d\Gamma/dx\,dy]_{F\bar F}^{\rm canon}$.
  }
\end {center}
\end {figure}

There are three large-$\Nc$ color routings of the $4\bar 4$ diagram,
shown in fig.\ \ref{fig:44route}.%
\footnote{
  Our fig.\ \ref{fig:44route} is adapted from fig.\ 14 of ref.\ \cite{4point}.
  See footnote \ref{foot:x1234warning} of the current
  paper concerning the difference in line numbering convention.
}
Again, we choose the ``canonical''
routing to be the one ordered (1234) according to (\ref{eq:x1234}).
The total $4\bar 4$ contribution can then be written%
\footnote{
  We've written (\ref{eq:44bar}) in a way that most easily tracks
  how fig.\ \ref{fig:44route} was drawn, which was adapted from
  ref.\ \cite{4point}.  However, one may alternatively relabel
  the $(1243)$ routing in fig.\ \ref{fig:44route} as $(1342)$,
  which is equivalent since the direction one circles the
  circumference of the cylinder does not matter.  Then the three
  routings can be seen to be cyclic permutations of $(234){=}(y,z,x)$
  and so of $(x,y,z)$.  If desired, that cyclic-permutation relationship
  could be made manifest by rewriting (\ref{eq:44bar}) as
  \[
  \left[ \frac{d\Gamma}{dx\,dy} \right]_{4\bar 4}
  =
  \left[ \frac{d\Gamma}{dx\,dy} \right]_{4\bar 4}^{\rm canon}
  +~
  [(x,y,z) \rightarrow (y,z,x)]
  ~+~
  [(x,y,z) \rightarrow (z,x,y)] .
  \]
}
\begin {equation}
  \left[ \frac{d\Gamma}{dx\,dy} \right]_{4\bar 4}
  =
  \left[ \frac{d\Gamma}{dx\,dy} \right]_{4\bar 4}^{\rm canon}
  +~
  [(x,y,z) \rightarrow (z,y,x)]
  ~+~
  [(x,y,z) \rightarrow (z,x,y)] ,
\label {eq:44bar}
\end {equation}
where here $x$, $y$, and $z \,{\equiv}\, 1{-}x{-}y$
represent the three daughters
of the $g\to ggg$ splitting process.  We now generalize to include
instantaneous diagrams by writing
\begin {equation}
  \left[ \frac{d\Gamma}{dx\,dy} \right]_{F\bar F}
  =
  \left[ \frac{d\Gamma}{dx\,dy} \right]_{F\bar F}^{\rm canon}
  +~
  [(x,y,z) \rightarrow (z,y,x)]
  ~+~
  [(x,y,z) \rightarrow (z,x,y)] .
\end {equation}
The diagrams which contribute to the canonical routing (1234) are shown in
fig.\ \ref{fig:FFcanon}.

For the color and helicity factors, the simplest diagrams are those
involving only longitudinally polarized gluon interactions, for which the
factors [see fig.\ \ref{fig:rules} and eqs.\ (\ref{eq:shorthand})] are
\begin {subequations}
\label {eq:FFfactors}
\begin {align}
  [I_{12} \, \bar I_{12}]_{\rm canon} &\propto
     \tfrac12 |i g^2 c_{12} h_{12} i_{12}|^2 ,
\\
  [I_{14} \, \bar I_{14}]_{\rm canon} &\propto
     \tfrac12 |i g^2 c_{14} h_{14} i_{14}|^2 ,
\\
  I_{12} \, \bar I_{14} &\propto
    (i g^2 c_{12} h_{12} i_{12})(i g^2 c_{14} h_{14} i_{14})^* ,
\\
  I_{14} \, \bar I_{12} &\propto
    (i g^2 c_{14} h_{14} i_{14})(i g^2 c_{12} h_{12} i_{12})^* ,
\end {align}
where the factors of $\tfrac12$ arise for diagrams that have
two color routings when only one of those two routings is included in
fig.\ \ref{fig:FFcanon}.
The diagrams involving the fundamental 4-gluon vertex are a little more
subtle, but we can again leverage previous results.
The color contractions and particle numbering
in the $[4 \, \bar I_{12}]_{\rm canon}$ diagram of fig.\ \ref{fig:FFcanon}
are identical to those for the $[4\bar y\bar x]_{\rm canon}$ diagram discussed
earlier.  So, the appropriate piece of the 4-gluon vertex that contributes
to this particular color routing will be the same as that quoted
in (\ref{eq:4prop1}), taken in turn from ref.\ \cite{4point}.
Combining with the factors for $I_{12}$ in the conjugate amplitude
then gives
\begin {equation}
  [4 \, \bar I_{12}]_{\rm canon} \propto
     \bigl\{
       -i g^2 \bigl[
         \tfrac12 c_{12}(h_{13}{-}h_{14}) + c_{14}(h_{12}{-}h_{13})
       \bigr]
     \bigr\} \,
     (i g^2 c_{12} h_{12} i_{12})^* .
\label {eq:4I12}
\end {equation}
Though maybe not at first obvious from the way the diagrams have been
drawn, the $[4 \, \bar I_{14}]_{\rm canon}$ diagram is the same as the
$[4 \, \bar I_{12}]_{\rm canon}$ diagram except for interchange of particles
$2$ and $4$ (i.e. $x\leftrightarrow y$).  To see that the color
routings are the same after that interchange, remember that it doesn't
matter whether one circles the cylinder one way and names the
routing (1234) or circles the other way and names it in reverse
order (1432).  All that matters in the large-$\Nc$ limit is which lines are
neighbors going around the cylinder.%
\footnote{
  \label{foot:integrated}
  It also doesn't matter that we conventionally draw some lines as
  continuing very slight beyond the last interaction vertex.
  Since we only compute $p_\perp$-integrated rates,
  the evolution of all daughters of the splitting process
  can be thought of as stopping the {\it instant}
  they are emitted in both the amplitude and conjugate amplitude.
  (See section 4.1 of ref.\ \cite{2brem}.)
}
So, by swapping particles 2 and 4 in (\ref{eq:4I12}) while remembering
that our definitions of $c_{1n}$ and $i_{1n}$ imply
$c_{12} \leftrightarrow -c_{14}$ and $i_{12} \leftrightarrow -i_{14}$
under such a swap,
\begin {equation}
  [4 \, \bar I_{14}]_{\rm canon} \propto
     \bigl\{
       -i g^2 \bigl[
         \tfrac12 c_{14}(h_{12}{-}h_{13}) + c_{12}(h_{13}{-}h_{14})
       \bigr]
     \bigr\} \,
     (i g^2 c_{14} h_{14} i_{14})^* .
\end {equation}
The color and helicity factors are insensitive to the time ordering
of the vertices, and so the factors for $[I_{12} \, \bar 4]_{\rm canon}$ and
$[I_{14} \, \bar 4]_{\rm canon}$ are just the complex conjugates of those
for $[4 \, \bar I_{12}]_{\rm canon}$ and $[4 \, \bar I_{14}]_{\rm canon}$:
\begin {align}
  [I_{12} \, \bar 4]_{\rm canon} \propto
     (i g^2 c_{12} h_{12} i_{12})
     \bigl\{
       -i g^2 \bigl[
         \tfrac12 c_{12}(h_{13}{-}h_{14}) + c_{14}(h_{12}{-}h_{13})
       \bigr]
     \bigr\}^* ,
\\
  [I_{14} \, \bar 4]_{\rm canon} \propto
     (i g^2 c_{14} h_{14} i_{14})
     \bigl\{
       -i g^2 \bigl[
         \tfrac12 c_{14}(h_{12}{-}h_{13}) + c_{12}(h_{13}{-}h_{14})
       \bigr]
     \bigr\}^* .
\end {align}
This
{\it overall}\/ complex conjugation doesn't actually make a difference,
since the above are real-valued.
Finally, there is the $4\bar 4$ diagram, which has the three color routings
shown in fig.\ \ref{fig:44route}.  As discussed in ref.\ \cite{4point},
the contribution of each color routing is just one third of
what the total would be if we naively ignored the necessity of
splitting the $4\bar 4$ diagram into different large-$\Nc$ color
routings. So,
\begin {equation}
  [4\,\bar 4]_{\rm canon} \propto
  \tfrac13 \Bigl|
    -ig^2 \bigl[
      c_{14}(h_{12}{-}h_{13}) + c_{12}(h_{13}{-}h_{14}) + c_{13}(h_{14}{-}h_{12})
    \bigr]
  \Bigr|^2 .
\end {equation}
\end {subequations}

Recall that we defined the $(c_{12},c_{13},c_{14})$
to cyclically permute under permutations of the indices $(234)$.
So (\ref{eq:ccOriginal}) gives
\begin {subequations}
\begin {equation}
   \langle c_{12} c_{12} \rangle = \langle c_{13} c_{13} \rangle
   = \langle c_{14} c_{14} \rangle = \CA^2 ,
\end {equation}
\begin {equation}
  \langle c_{12} c_{13} \rangle = \langle c_{13} c_{14} \rangle
  = \langle c_{14} c_{12} \rangle = -\tfrac12 \CA^2 .
\end {equation}
\end {subequations}
From the definition (\ref{eq:h1n})
of the $h_{1n}$, final/initial helicity summing/averaging gives%
\footnote{
  There are no ultraviolet divergences associated with the time-ordered
  diagrams in this paper.  We will not need to use dimensional regularization
  (which was used for other diagrams in refs.\ \cite{seq,dimreg,QEDnf,qcd}),
  and so the number of possible ``helicities'' is simply 2 in this paper.
}
\begin {subequations}
\begin {equation}
   \langle h_{12} h_{12} \rangle = \langle h_{13} h_{13} \rangle
   = \langle h_{14} h_{14} \rangle = 2 ,
\end {equation}
\begin {equation}
  \langle h_{12} h_{13} \rangle = \langle h_{13} h_{14} \rangle
  = \langle h_{14} h_{12} \rangle =  1 .
\end {equation}
\end {subequations}
Eqs.\ (\ref{eq:FFfactors}) then yield
(after absorbing a common factor of $\CA^2 g^4$ into the proportionality)
\begin {align}
  [I_{12} \, \bar I_{12}]_{\rm canon} &\propto i_{12}^2 ,
\\
  [I_{14} \, \bar I_{14}]_{\rm canon} &\propto i_{14}^2 ,
\\
  I_{12} \, \bar I_{14} &\propto -\tfrac12 i_{12} i_{14} ,
\\
  I_{14} \, \bar I_{12} &\propto -\tfrac12 i_{12} i_{14} ,
\\
  [4 \, \bar I_{12}]_{\rm canon} = [I_{12} \, \bar 4]_{\rm canon}
      &\propto \tfrac12 i_{12} ,
\\
  [4 \, \bar I_{14}]_{\rm canon} = [I_{14} \, \bar 4]_{\rm canon}
      &\propto -\tfrac12 i_{14} ,
\\
  [4\,\bar 4]_{\rm canon} &\propto 3 .
\end {align}
Adding all nine color-routed diagrams of fig.\ \ref{fig:FFcanon} together,
\begin {equation}
  [F\,\bar F]_{\rm canon} \propto
    3 + i_{12}^2 + i_{14}^2 - i_{12} i_{14} + i_{12} - i_{14} ,
\end {equation}
to be compared with just
$[4\,\bar 4]_{\rm canon} \propto 3$.
So, we can convert the result for $4\bar 4$ in ref.\ \cite{4point} to
the more general result for $F\bar F$ by
\begin {equation}
  \left[ \frac{d\Gamma}{dx\,dy} \right]_{F\bar F}^{\rm canon}
  =
  \bigl[
     1 +
     \tfrac13 ( i_{12}^2 + i_{14}^2 - i_{12} i_{14} + i_{12} - i_{14} )
  \bigr]
  \left[ \frac{d\Gamma}{dx\,dy} \right]_{4\bar 4}^{\rm canon} .
\label {eq:convert44}
\end {equation}
A summary of the final rate formula is given in appendix \ref{app:summary}.

% ============================================================================

\section {Virtual corrections to $g{\to}gg$ with 4-gluon interactions}
\label {sec:virt}

\subsection {Basic results}
\label {sec:virtBasic}

\begin {figure}[t]
\begin {center}
  \includegraphics[scale=0.45]{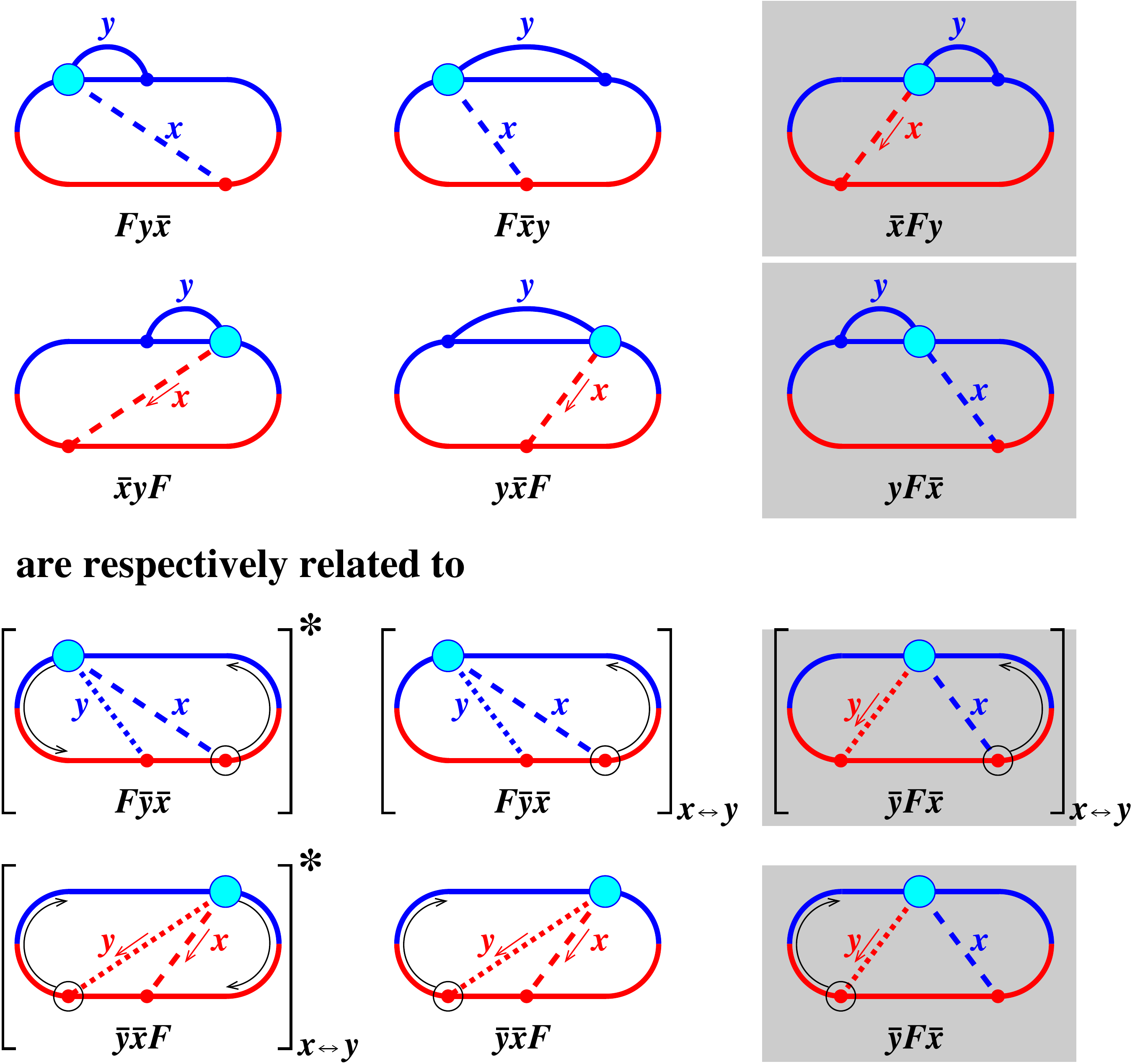}
  \caption{
     \label{fig:Transform}
     Relation of the virtual diagrams of figs.\ \ref{fig:FdiagsVirtI}
     and \ref{fig:FdiagsVirtII} to the real $g{\to}ggg$ diagrams of
     fig.\ \ref{fig:FdiagsReal} by various combinations of
     front- and back-end transformations, complex conjugation, and
     swapping variable names $x\leftrightarrow y$.
     Complex conjugation corresponds to flipping a diagram upside down
     and swapping the colors blue (amplitude) and red (conjugate
     amplitude).  Complex conjugation is irrelevant, however, because
     ultimately we want to take $2\Re(\cdots)$ of all the virtual diagrams,
     i.e.\ add them to their complex conjugates.
     Shading above indicates diagrams that happen to be exactly zero,
     because the $\bar yF\bar x$ diagram vanishes for the reason
     described in section \ref{sec:yFx}.
  }
\end {center}
\end {figure}

In previous work \cite{QEDnf,qcd}, we showed how almost all of the diagrams
considered
there for virtual corrections to single splitting $g{\to}gg$ could be simply
related to diagrams for real double splitting $g{\to}ggg$ through what
we call front- and/or back-end transformations.
Those same techniques can be applied to all of the virtual diagrams
of this paper.
In particular, fig.\ \ref{fig:Transform} depicts
graphically how the virtual diagrams of figs.\ \ref{fig:FdiagsVirtI} and
\ref{fig:FdiagsVirtII} are related to the $g{\to}ggg$ diagrams of
fig.\ \ref{fig:FdiagsReal}.
The front- and back-end transformations are represented by the black
arrows in the bottom half of fig.\ \ref{fig:Transform}.
Graphically,
front-end transformations correspond to sliding the earliest-time vertex
in the interference
diagram around the front end of the diagram from amplitude to
conjugate amplitude or vice versa.  Back-end transformation
correspond to similarly sliding the
latest-time vertex around the back end of the diagram.
The transformations depicted by fig.\ \ref{fig:Transform} involve
various combinations, as indicated,
of front-end transformations, back-end transformations,
complex conjugation, and swapping variable names $x\leftrightarrow y$.

The simplest transformation is a back-end transformation, where the
latest-time vertex changes sign because a $-i\,\delta H$ perturbation
in the amplitude
(from perturbing the evolution operator $e^{-i H t}$)
moves to become a $+i\,\delta H$ perturbation
in the conjugate amplitude
(from perturbing $(e^{-i H t})^* = e^{+i H t}$),
or vice versa.
So fig.\ \ref{fig:Transform} tells us that%
\footnote{
  See section 4.1 of ref.\ \cite{QEDnf} and section 2.2 of ref.\ \cite{qcd}
  for earlier discussion and application of back-end transformations.
}
\begin {equation}
   \left[ \frac{d\Gamma}{dx} \right]_{F\bar x y}
   =
   - \tfrac12
   \int_0^{1-x} dy \>
   \left\{
       \left[ \frac{d\Gamma}{dx\,dy} \right]_{F\bar y\bar x}
       ~\mbox{with}~ x \leftrightarrow y
   \right\},
\label {eq:FXby}
\end {equation}
where the loop momentum fraction $y$ has been integrated over.%
\footnote{
  In LCPT, the lightcone momentum variables $p^+$ for transversely
  polarized gluons must all be positive, whether real or virtual.
  This restricts the integration of $y$ to $0 < y < 1{-}x$ for
  the diagrams of fig.\ \ref{fig:FdiagsVirtI} and to
  $0 < y < 1$ for those of fig.\ \ref{fig:FdiagsVirtII}.
}
The overall factor of $\tfrac12$ is the symmetry factor of the (blue) loop in
the amplitude of the $F\bar x y$ diagram in fig.\ \ref{fig:Transform}.

Front-end transformations are similar, but the momentum fractions of
the lines must be adjusted since they are defined relative to the
parent energy $E$ of the entire splitting process, and which line is
the parent changes under a front-end transformation.
For the case $y\bar x F$ in fig.\ \ref{fig:Transform},
where a $y$-emission 3-gluon vertex is being slid
around the front of the diagram, this is%
\footnote{
  See section 4.2 of ref.\ \cite{QEDnf} or section 2.2 of ref.\ \cite{qcd},
  but exchange the label $x$ for $y$ there.
  Also, one does not need the factors of
  $(1{-}x)^{-\eps}$ or $(1{-}y)^{-\eps}$ that accompany front-end transformations
  in that discussion because our diagrams here have no ultraviolet divergences
  and do not require dimensional regularization.
}
\begin {equation}
   \left[ \frac{d\Gamma}{dx} \right]_{y\bar xF}
   =
   -
   \tfrac12 \int_0^1 dy \>
   \left\{
     \left[ \frac{d\Gamma}{dx\,dy} \right]_{\bar y\bar x F}
     ~\mbox{with}~
     (x,y,E) \longrightarrow
     \Bigl( \frac{x}{1{-}y} \,,\, \frac{{-}y}{1{-}y} \,,\, (1{-}y)E \Bigr)
   \right\} .
\label {eq:yXbF}
\end {equation}
The sign change appearing in the transformation $y \to -y/(1{-}y)$
arises because our
(very useful) convention \cite{2brem}
is that particles in time-ordered interference
diagrams have positive or negative
momentum fractions depending on whether they are emitted first
in the amplitude (blue lines) or conjugate amplitude (red lines),
respectively.%
\footnote{
  Because of these sign changes,
  it was necessary in refs.\ \cite{QEDnf,qcd} to
  add absolute value signs appropriately to expressions involving
  DGLAP splitting functions, such as in eq.\ (A.30) of ref.\ \cite{QEDnf}
  and eqs.\ (A.5) and (A.23) of
  ref.\ \cite{qcd}, so that the expressions for combinations of
  splitting functions for a diagram remained correct after front-end
  transformation.  The analogous factors in this paper that arise from
  DGLAP splitting functions are the
  $(\zeta_{12},\zeta_{13},\zeta_{14})$ of (\ref{eq:zetaij}), and thence
  the $\zeta_{(4)}$ of (\ref{eq:zeta4alt}).  The $1{-}y$ factors in
  those equations arise from the longitudinal momentum fraction of
  the intermediate line in the $F\bar y\bar x$ diagram, and the other
  factors of $x$, $y$, and $z$ arise from the momentum fractions of
  the three final-state daughters.  One could make this expression
  safe for any type of front-end transformation by replacing
  $x$,$y$,$z$, and $1{-}y$ by $|x|$, $|y|$, $|z|$, and $|1{-}y|$
  respectively.  The first three replacements make no difference to
  the expression, and $1{-}y \rightarrow |1{-}y|$ will not matter because
  all of the front-end transformations we will use keep $1{-}y$ positive.
}

For the $\bar x y F$ diagram in fig.\ \ref{fig:Transform}, we combine
the above transformation with a back-end transformation, conjugation,
and $x\leftrightarrow y$:
\begin {align}
   \left[ \frac{d\Gamma}{dx} \right]_{\bar x y F}
   &=
   + \tfrac12 \int_0^1 dy \>
   \Biggl\{
     \left[ \frac{d\Gamma}{dx\,dy} \right]_{\bar y\bar x F}
     ~\mbox{with}~
     (x,y,E) \longrightarrow
     \Bigl( \frac{x}{1{-}y} \,,\, \frac{{-}y}{1{-}y} \,,\, (1{-}y)E \Bigr)
\nonumber\\ & \hspace{12em}
     ~\mbox{followed by}~ x \leftrightarrow y
   \Biggr\}^*
\nonumber\\
   &=
   + \tfrac12 \int_0^1 dy \>
   \left\{
     \left[ \frac{d\Gamma}{dx\,dy} \right]_{\bar y\bar x F}
     ~\mbox{with}~
     (x,y,E) \longrightarrow
     \Bigl( \frac{y}{1{-}x} \,,\, \frac{{-}x}{1{-}x} \,,\, (1{-}x)E \Bigr)
   \right\}^* .
\label {eq:XbyF}
\end {align}

For the case $Fy\bar x$,
where the front-end transformation is of a 4-gluon interaction,
the momentum fraction transformations are correspondingly different
because the particle line that becomes the new parent is different,
and also because the front-end transformation moves
two emissions ($x$ and $y$) from amplitude to conjugate amplitude:%
\footnote{
  See section 4.2 of ref.\ \cite{QEDnf}, and in particular eq.\ (4.5)
  of that reference.
  The $x$ and $y$ in our $Fy\bar x$ diagram here correspond to the labels
  $y_e$ and $x_e$ there, respectively.
}
\begin {multline}
   \left[ \frac{d\Gamma}{dx\,dy} \right]_{F y\bar x}
   =
\\
   + \tfrac12 \int_0^{1-x} dy \>
   \left\{
     \left[ \frac{d\Gamma}{dx\,dy} \right]_{F\bar y\bar x}
     ~\mbox{with}~
     (x,y,E) \longrightarrow
     \Bigl(
       \frac{{-}x}{1{-}x{-}y} \,,\, \frac{{-}y}{1{-}x{-}y} \,,\, (1{-}x{-}y)E
     \Bigr)
   \right\}^* .
\label {eq:FyXb}
\end {multline}

% ---------------------------------------------------------------------------

\subsection {Integrable infrared divergence from instantaneous interactions}
\label {sec:PV}

\begin {figure}[t]
\begin {center}
  \includegraphics[scale=0.45]{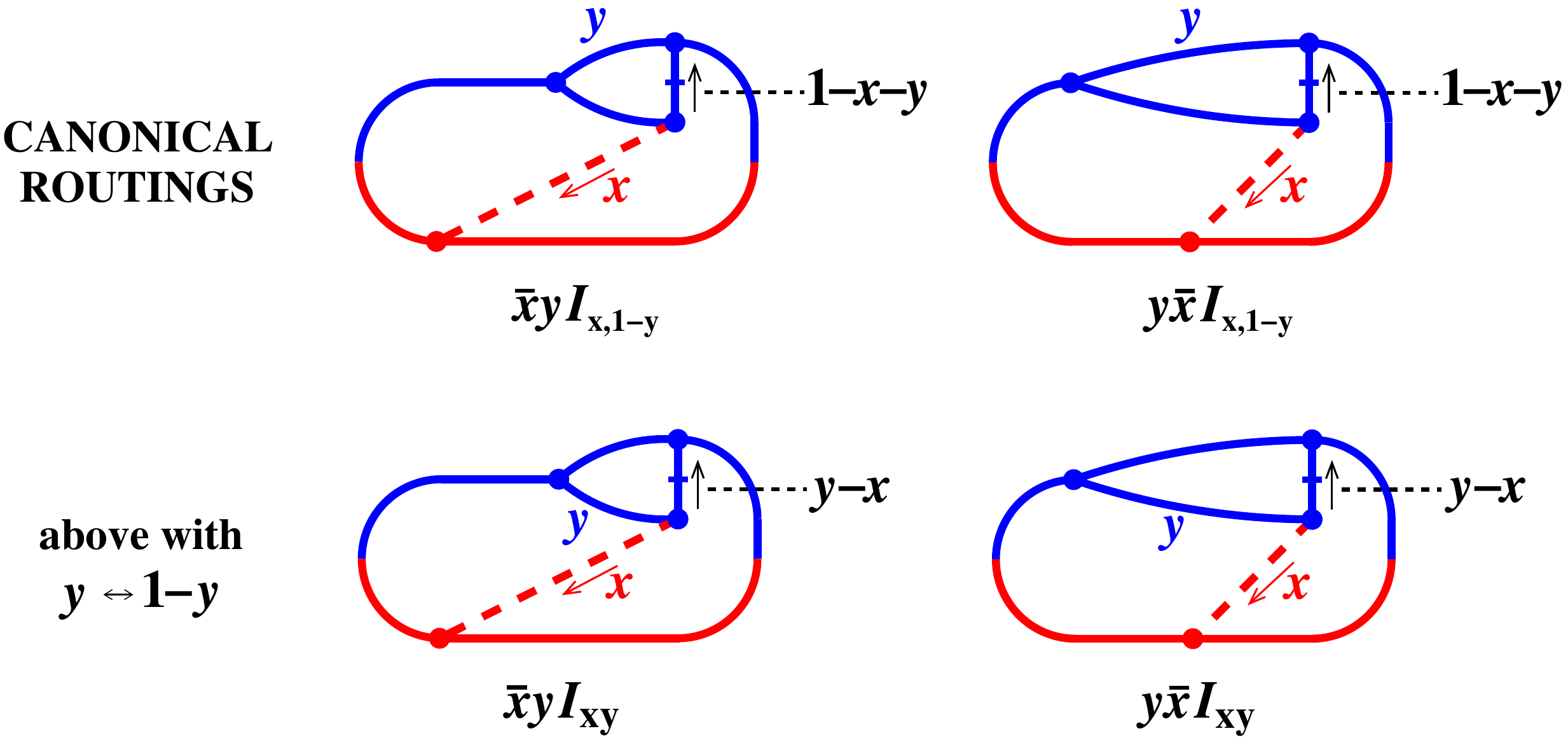}
  \caption{
     \label{fig:div}
     Diagrams with infrared divergences associated with
     longitudinally polarized gluon exchange.  The divergences
     occur at either $y \to 1{-}x$ (top line) or
     $y \to x$ (bottom line).
  }
\end {center}
\end {figure}

Of all the various diagrams represented by
figs.\ \ref{fig:FdiagsReal}--\ref{fig:FdiagsVirtII}, there are
four particular cases where divergences arise because the
$q^+$ of an exchanged longitudinal gluon may become zero.
Those cases are shown in fig.\ \ref{fig:div} and are all virtual
diagrams corresponding to certain types of rescattering corrections to a
leading-order single splitting $g {\to} gg$.
The loop momentum fraction $y$ is integrated over $0 < y < 1$ in these
diagrams, and the divergences occur at $y = 1{-}x$ for the two diagrams
in the top line of fig.\ \ref{fig:div} and at $y = x$ for the
other two diagrams.

In order to reduce the number of things to think about, we may focus on
just the top line of fig.\ \ref{fig:div}.
These are the divergent contributions
from $\bar x y F$ and $y\bar x F$ 
that are obtained by applying the relevant transformations
(fig.\ \ref{fig:Transform})
to only the {\it canonical}\/ color routing of $\bar y\bar x F$.
The other color routing of $\bar y\bar x F$ corresponds to swapping
$x \leftrightarrow z$, which, after transformation, corresponds to
swapping $y \leftrightarrow 1{-}y$ in $\bar x y F$ and $y\bar x F$.%
\footnote{
  For the transformation in (\ref{eq:XbyF}) that gives $\bar x y F$,
  it is easy to check algebraically that
  $x\leftrightarrow z$ transforms to $y \leftrightarrow 1{-}y$.
  For the transformation in (\ref{eq:yXbF}) that gives $y\bar x F$,
  $x\leftrightarrow z$ instead transforms to $x \leftrightarrow 1{-}x$.
  However, $x \leftrightarrow 1{-}x$ on either
  $y\bar x I$ diagram in fig.\ \ref{fig:Transform}
  results in the same diagram that $y \leftrightarrow 1{-}y$ does, even without
  yet integrating over $y$.
  (See also footnote \ref{foot:integrated}.)
}
Since we are integrating $y$ over $0 < y < 1$ in these particular
virtual diagrams, adding in the contribution from swapping
$y \leftrightarrow 1{-}y$ is equivalent to multiplying the integral
of the canonical routing by a factor of 2.
So, we will rewrite (\ref{eq:yXbF}) and (\ref{eq:XbyF}) as
\begin {subequations}
\label {eq:div1}
\begin {align}
   \left[ \frac{d\Gamma}{dx} \right]_{y\bar xF}
   &=
   -
   \int_0^1 dy \>
   \left\{
     \left[ \frac{d\Gamma}{dx\,dy} \right]_{\bar y\bar x F}^{\rm canon}
     ~\mbox{with}~
     (x,y,E) \longrightarrow
     \Bigl( \frac{x}{1{-}y} \,,\, \frac{{-}y}{1{-}y} \,,\, (1{-}y)E \Bigr)
   \right\} ,
\label {eq:div1a}
\\
   \left[ \frac{d\Gamma}{dx} \right]_{\bar x y F}
   &=
   + \int_0^1 dy \>
   \left\{
     \left[ \frac{d\Gamma}{dx\,dy} \right]_{\bar y\bar x F}^{\rm canon}
     ~\mbox{with}~
     (x,y,E) \longrightarrow
     \Bigl( \frac{y}{1{-}x} \,,\, \frac{{-}x}{1{-}x} \,,\, (1{-}x)E \Bigr)
   \right\}^* .
\label {eq:div1b}
\end {align}
\end {subequations}
Comparing to the earlier versions, notice the restriction
``canon'' now on the $\bar y\bar x F$ rates,
and correspondingly the removal of the overall factors of $\tfrac12$.
The only divergences in the $y$ integration are now the ones at
$y = 1{-}x$, from the top line of fig.\ \ref{fig:div}.
Individually, each of the two diagrams in (\ref{eq:div1})
has a $1/(1{-}x{-}y)^2$ divergence
as $y \to 1{-}x$ because of the $1/(q^+)^2 = 1/(p_1^+ + p_2^+)^2$
in fig.\ \ref{fig:rules} associated with the
propagator of the longitudinally polarized gluon.

In what follows, it will be convenient to get rid of the complex
conjugation in (\ref{eq:div1b}) by noting that ultimately these
diagrams must be added to their complex conjugates, as noted at
the bottom of fig.\ \ref{fig:FdiagsVirtII}.  It will also be
convenient to add together all the diagrams (including the conjugates) of
fig.\ \ref{fig:FdiagsVirtII}.  These diagrams represent the
4-gluon interaction contributions to a class of diagrams that
were called ``Class II'' virtual diagrams in ref.\ \cite{qcd},
and we adopt that nomenclature here for the sum.
Remembering that the shaded diagram in fig.\ \ref{fig:FdiagsVirtII}
is zero, we then have
\begin {subequations}
\label {eq:classIIcanon}
\begin {equation}
  \left[ \frac{d\Gamma}{dx} \right]_{\rm F,~virt~II}
  =
  2\Re \left\{
    \left[ \frac{d\Gamma}{dx} \right]_{y\bar x F}
    +
    \left[ \frac{d\Gamma}{dx} \right]_{\bar x y F}
  \right\}
  =
  2\Re \int_0^1 dy \> {\cal F}(x,y)
\label {eq:classII}
\end {equation}
with
\begin {align}
  {\cal F}(x,y) \equiv &
   -
   \left\{
     \left[ \frac{d\Gamma}{dx\,dy} \right]_{\bar y\bar x F}^{\rm canon}
     ~\mbox{with}~
     (x,y,E) \longrightarrow
     \Bigl( \frac{x}{1{-}y} \,,\, \frac{{-}y}{1{-}y} \,,\, (1{-}y)E \Bigr)
   \right\}
\nonumber\\ &
   + \left\{
     \left[ \frac{d\Gamma}{dx\,dy} \right]_{\bar y\bar x F}^{\rm canon}
     ~\mbox{with}~
     (x,y,E) \longrightarrow
     \Bigl( \frac{y}{1{-}x} \,,\, \frac{{-}x}{1{-}x} \,,\, (1{-}x)E \Bigr)
   \right\} .
\label {eq:calF}
\end {align}
\end {subequations}
Now that we've added the diagrams together and avoided any complex
conjugation in (\ref{eq:calF}), it turns out that the
$1/(1{-}x{-}y)^2$ divergences of the two terms cancel,
leaving behind a milder $1/(1{-}x{-}y)$ divergence.%
\footnote{
  Here's one way to see the cancellation without drilling down into
  the specific formula for
  $[d\Gamma/dx\,dy]^{\rm canon}_{\bar y\bar x F}$.
  First, note that the two diagrams on the top line of fig.\ \ref{fig:div}
  are topologically unchanged if we simultaneously replace both
  $y \to 1{-} y$ (and so swap the two entirely-blue lines in the amplitude)
  and $x \to 1{-} x$ (and so interchange the two daughter lines in the
  diagram).  Moreover, if the diagrams are drawn on the cylinder to
  emphasize their color routing, these changes preserve the color routing:
  lines that were neighbors going around the cylinder remain neighbors after
  the change.  So (\ref{eq:div1a}) would have given the same result with
  the alternate substitute rule
  $
     (x,y,E) \longrightarrow
     \Bigl( \frac{1{-}x}{y} \,,\, \frac{{-}(1{-}y)}{y} \,,\, yE \Bigr)
  $.
  In the limit $y \to 1{-}x$, both this and the rule in (\ref{eq:div1b})
  for the other diagram give the same substitution
  $(x,y,E) \rightarrow \bigl(1, -\frac{x}{1-x}, (1{-}x)E \bigr)$.
  That is, the differences are suppressed by $O(1{-}x{-}y)$.
  That means that both diagrams give the same contribution to
  (\ref{eq:calF}) in the $y \to 1{-}x$
  limit {\it except}\/ for the overall sign difference there,
  and so they cancel,
  up to corrections suppressed by one relative power of
  $1{-}x{-}y$.  We have verified numerically that the subleading
  $1/(1{-}x{-}y)$ divergence of these diagrams does not cancel.
}
To make our discussion more compact, we'll loosely refer to this as
a $1/z$ divergence with $z \equiv 1{-}x{-}y$.  However, unlike
the discussion of $g{\to}ggg$ processes in section \ref{sec:gTOggg},
$z$ is {\it not}\/ the momentum fraction of any final-state
daughter of the single splitting processes being considered here,
and $z$ need not be positive.

The nice thing about a $1/z$ divergence is that, since the
integral $\int_0^1 dy$ associated with the loop integrals
of fig.\ \ref{fig:div} span both signs of $z=1{-}x{-}y$,
the integral $\int dz/z$ will be finite: the divergent contributions
from $z$ slightly negative will cancel those from $z$ slightly
positive.  Though the answer will
be finite, the subtlety lies in figuring out what finite piece
will be left over.
We will address this first formally, and then
as a practical matter to allow the $y$ integration in (\ref{eq:classII}) to
be performed numerically in applications.

% ..........................................................................

\subsubsection {Disambiguation}

Other diagrams in previous work \cite{qcd} (which did not include
longitudinally-polarized gluon exchange) had infrared divergences
associated with one of the transversely polarized gluons becoming soft.
There, we regulated those divergences by introducing a small infrared
cut-off $\delta\ll 1$ on all momentum fractions such as $x$, $y$, $1{-}y$,
etc.  That's equivalent to saying that we insisted that
$p^+ > (p^+)_{\rm min} \equiv P^+ \delta$, where $P$ is the momentum of
the initial particle in the overlapping splitting process.
In LCPT, the transversely polarized gluons all propagate forward
in time with $p^+ > 0$.
But there is no restriction on the longitudinally-polarized gluons,
which have been integrated out and for which there is no forward
direction of light-cone time since they mediate instantaneous
interactions.  Their $q^+$ can have either sign.  One can regulate
the {\it magnitude} of $q^+$ in a way consistent with
the transversely polarized gluons:
$|q^+| > (p^+)_{\rm min} = P^+ \delta$.
Given that the infrared regulator $\delta$ is to be formally chosen
as arbitrarily small, that's equivalent to regulating our net
$1/z$ divergence with a principal value (also known as principal part)
prescription:
\begin {equation}
   \PV \Bigl[ \frac{1}{z} \Bigr]
   =
   \frac{\theta(|z|-\delta)}{z} \,,
\label {eq:PVdef1}
\end {equation}
where $\theta$ is the unit step function.
In terms of $i\eps$ prescriptions, the principal value (PV)
can alternatively be defined as
\begin {equation}
   \PV \Bigl[ \frac{1}{z} \Bigr]
   =
   \frac12 \left( \frac{1}{z-i\eps} + \frac{1}{z+i\eps} \right)
   =
   \frac{z}{z^2+\eps^2} \,.
\label {eq:PVdef2}
\end {equation}
Both (\ref{eq:PVdef1}) and (\ref{eq:PVdef2}) cut off small values
of $z$ while keeping $\PV[1/z]$ real valued.  The only difference is
that one is a sharp IR cut-off on $|z|$ while the other is smoothed out.%
\footnote{
  If $f(z)$ is any function that is smooth at $z{=}0$, then both
  prescriptions give the same answer for integrating $\PV[1/z]\,f(z)$
  across $z{=}0$.
  If desired,
  they can also be made to give exactly the same
  (infrared regulated) answer for
  integrating $(\PV[1/z])^2 \, f(z)$ --- an integral that gives
  $2 f(0)/\delta$ plus a finite piece as $\delta\to 0$
  --- by choosing $\delta = 4\eps/\pi$.
}

The use of principle value prescriptions for such
$1/q^+$ divergences in lightcone gauge had a convoluted early history
\cite{Leibbrandt}.
Here, we rely on the more recent analysis by Chirilli, Kovchegov
and Wertepny \cite{CKW,CKWolder}, which shows how various
$i\eps$ prescriptions for $1/q^+$ divergences in lightcone gauge can
be understood as corresponding to different sub-gauge choices of
lightcone gauge
and correspondingly to different choices of boundary conditions for gauge
fields as $x^- \to \pm\infty$.
Sub-gauges arise because the lightcone gauge condition $A^+{=}0$
does not by itself uniquely determine the gauge.
In what they call PV sub-gauge, the
Feynman propagator is
\begin {subequations}
\label {eq:GPV}
\begin {equation}
   G^{\mu\nu}(q) =
   \frac{i}{q^2+i\eps} \, \Delta^{\mu\nu}(q)
\end {equation}
with
\begin {equation}
   \Delta^{\mu\nu}(q) =
   - \biggl\{
     g^{\mu\nu} - (q^\mu n^\nu + q^\nu n^\mu) \PV\Bigl[ \frac{1}{q\cdot n} \Bigr]
   \biggr\} .
\label {eq:Delta}
\end{equation}
\end {subequations}
They also explicitly check in certain
examples the equivalence of calculations performed in different sub-gauges,
one of which is PV sub-gauge.

For LCPT and for our calculation, we want to separate the transverse and
longitudinal polarizations.
Algebraically manipulating (\ref{eq:GPV}) into the form of (\ref{eq:GLTdecomp})
while maintaining the prescriptions gives%
\footnote{
   In comparison to eqs.\ (12), (16) and (17) of ref.\ \cite{CKWolder},
   our $\Delta^{\mu\nu}$ is their $-D^{\mu\nu}$.
}
\begin {subequations}
\label {eq:DeltaDecomp}
\begin {equation}
   \Delta^{\mu\nu}(q) = \Delta_{\rm T}^{\mu\nu}(q) + \Delta_{\rm L}^{\mu\nu}(q)
\end {equation}
with
\begin {equation}
    \Delta^{\mu\nu}_{\rm T}(q) =
    \sum_\lambda \eps_{(\lambda)}^\mu(q) \, \eps_{(\lambda)}^{\nu*}(q) ,
    \qquad
    \Delta^{\mu\nu}_{\rm L}(q) =
    n^\mu n^\nu q^2 \biggl( \PV\Bigl[ \frac{1}{q\cdot n} \Bigr] \biggr)^2
\label {eq:DeltaLTPV}
\end {equation}
and
\begin {equation}
   (\eps^+,\eps^-,\beps)_{(\lambda)}
   =
   \Bigl( 0, \beps_{(\lambda)}\cdot\q \PV\Bigl[ \frac{1}{q^+} \Bigr],
          \beps_{(\lambda)} \Bigr) ,
\end {equation}
\end {subequations}
This reproduces a prescription proposed earlier by
Zhang and Harindranath \cite{ZHboundary} in the context of LCPT.%
\footnote{
  See in particular eqs.\ (21) and (22) of \cite{ZHboundary}
  and the discussion following them.  A technical point is that
  Zhang and Harindranath take the boundary condition for the
  $(A^1,A^2)$ components of the gauge field to be
  ${\bm A}_\perp(x^- = +\infty) = - {\bm A}_\perp(x^-=-\infty)$,
  whereas Chirilli et al.\ \cite{CKW} find that the PV sub-gauge condition
  should be the slightly more general one that
  $\grad_\perp\cdot{\bm A}_\perp(x^- = +\infty)
   = - \grad_\perp\cdot{\bm A}_\perp(x^-=-\infty)$.
}

\begin {figure}[t]
\begin {center}
  \includegraphics[scale=0.45]{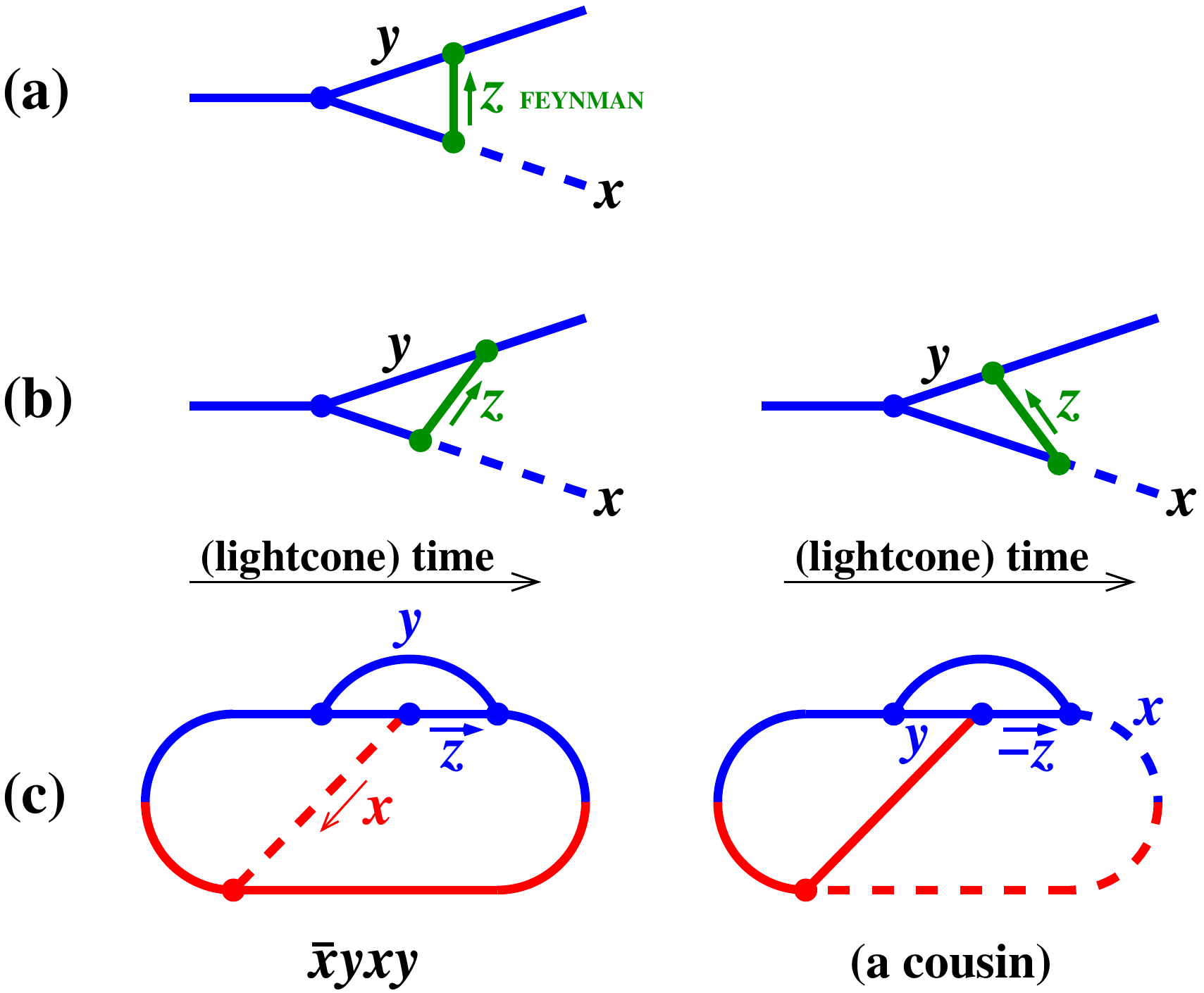}
  \caption{
     \label{fig:PVexample}
     (a) An ordinary Feynman diagram represent a one-loop vertex
     correction to single splitting $g{\to}gg$, where we have
     highlighted the line $z = 1{-}x{-}y$ in green.
     (b) The two time orderings in LCPT for the
     transverse-polarization contributions to that Feynman diagram.
     (c) Examples of rate diagrams that involve those time orderings
     of the amplitude.  (Other examples just
     correspond to different choices of how to time-order the
     splitting
     in the conjugate
     amplitude relative to the interactions in the amplitude.)
     The first diagram in (c), $\bar x y x y$, is an example of a Class I
     virtual correction analyzed in ref.\ \cite{qcd}.  There, Class I diagrams
     must be added to their ``cousins'' obtained by swapping the two daughters
     of the splitting,
     $x \to 1{-}x$.  The second diagram above is the cousin of the first,
     which can be seen by relabeling the loop variable $y$
     in the second diagram by $y \to 1{-}y$.  We haven't relabeled $y$ here
     because that would destroy the correspondence of the label $z=1{-}x{-}y$
     in (c) with the labeling of the original Feynman diagram in (a).
  }
\end {center}
\end {figure}

Let's now see a little more explicitly that our previous calculations
\cite{qcd} involving only transversely polarized gluons corresponded
implicitly to PV sub-gauge for Feynman propagators, and so the
longitudinally polarized gluon propagators in our current analysis
should be evaluated with the PV prescription as well.
Fig.\ \ref{fig:PVexample}a show an ordinary {\it Feynman} diagram
for the one-loop vertex correction to the amplitude for single splitting.
In keeping with the rest of this paper, we label lines by their
momentum fractions associated with $p^+$.  One of the lines is
labeled $y$, which we can take as our loop integration variable.
The line highlighted by being drawn in green is then $z = 1{-}x{-}y$.
Feynman diagrams implicitly contain all possible time orderings of
the interaction vertices, examples of which are shown in
fig.\ \ref{fig:PVexample}b.
In light-cone perturbation theory, time-orderings evaluate to zero
if any transversely-polarized gluon (whether real or virtual) has
a negative value of $p^+$ flowing forward in time.
If we focus on the part of the original Feynman
diagram of fig.\ \ref{fig:PVexample}a that comes only from
transverse polarizations, then fig.\ \ref{fig:PVexample}b is
a complete list of the corresponding time-orderings in LCPT.
The first time-ordering requires $0<x<1$ and $0 < y < 1{-}x$, which gives
$z = 1{-}x{-}y > 0$.
The second time-ordering requires $0<x<1$ and $1{-}x < y < 1$, which gives
$z = 1{-}x{-}y < 0$.  That's okay because $z$ flows backward in
time for that diagram, and it is $-z > 0$ that flows forward
in time.
Fig.\ \ref{fig:PVexample}c gives two examples of time-ordered rate
diagrams with the time orderings of fig.\ \ref{fig:PVexample}b in
the amplitude.
In the analysis of ref.\ \cite{qcd}, we took the conventional choice
in LCPT of regulating the infrared by
requiring all internal and external momentum fractions,
defined as flowing forward in time,
to be larger than some infrared regulator $\delta$.
This corresponds to $z > \delta$ for the first diagram in
fig.\ \ref{fig:PVexample}c and $-z > \delta$ for the second.
Taken together, that corresponds to using the infrared regulator
$|z| > \delta$ for the $z$ line in the original Feynman diagram
of fig.\ \ref{fig:PVexample}a.  By definition, that is regularization
of $z \to 0$ with a PV prescription (\ref{eq:PVdef1}) and so
corresponds to
working in PV sub-gauge.  But then longitudinal polarizations will
also be regulated with a PV prescription, as in
(\ref{eq:DeltaDecomp}).%
\footnote{
  There is a caveat to this discussion.  The propagator in
  (\ref{eq:DeltaDecomp}) is a vacuum propagator, which does not
  include medium effects.  So the lessons about consistent IR
  regularization drawn from
  fig.\ \ref{fig:PVexample} reflect a qualitative argument
  rather than a precise one.
}

% ..........................................................................

\subsubsection {Practical considerations}
\label{sec:average}

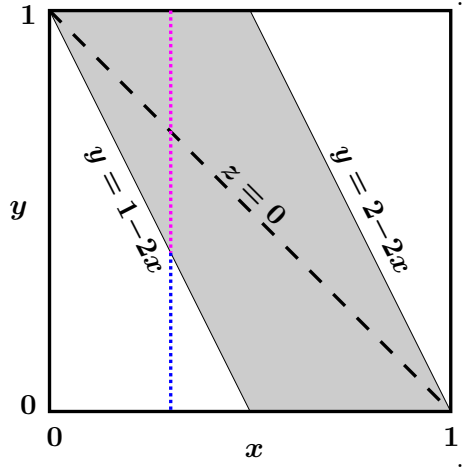
\begin {figure}[t]
\begin {center}
  \begin{picture}(175,175)(0,0)
    \put(20,20){\includegraphics[scale=0.35]{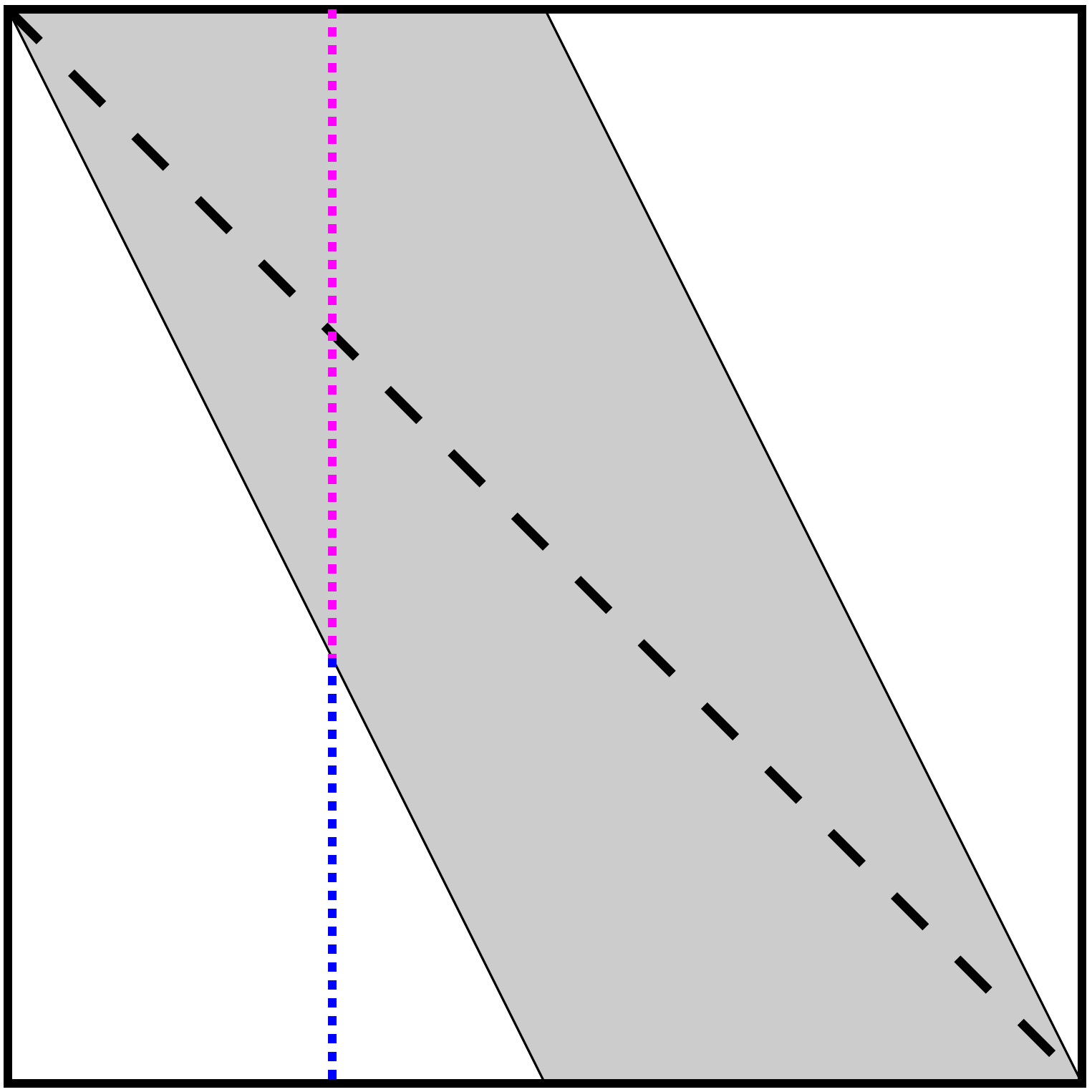}}
    % x axis
    \put(20,8){\textbf{0}}
    \put(170,8){\textbf{1}}
    \put(95,4){$\bm x$}
    % y axis
    \put(10,20){\textbf{0}}
    \put(10,168){\textbf{1}}
    \put(6,96){$\bm y$}
    % z=0 line
    \put(82,112){ \rotatebox{-45}{\boldmath$z=0$} }
    % shaded region boundaries
    \put(30,120){ \rotatebox{-63}{\boldmath$y=1{-}2x$} }
    \put(123,120){ \rotatebox{-63}{\boldmath$y=2{-}2x$} }
  \put(0,0){.}
  \put(0,175){.}
  \put(175,0){.}
  \put(175,175){.}
  \end{picture}
  \caption{
     \label {fig:PV}
     The shading shows, for each value of $x$,
     the maximally-sized interval of $y$ integration that is
     symmetric under $(x,z) \to (x,-z)$, which translates to
     $y \to 2(1{-}x)-y$.  The vertical dotted line is an example
     of the $y$ points for a particular value of $x$ and serves
     as a visual aid for the fact that, within the shaded region,
     the $y$ range with $z < 0$ has the same size as the
     $y$ range with $z > 0$.
  }
\end {center}
\end {figure}

Neither (\ref{eq:PVdef1}) nor (\ref{eq:PVdef2}) is
convenient for numerical integration, especially since the detailed
formula for ${\cal F}(x,y)$ is
complicated enough to be mildly expensive to evaluate numerically.
But now note
that $\PV(1/z)$ is odd in $z {\to}-z$.  Imagine that
we changed the integration
variables for $y$ to $z=1{-}x{-}y$ in (\ref{eq:classII})
to get an integral of the form
\begin {equation}
  \int dz \> f(z) \, \PV\Bigl[ \frac{1}{z} \Bigr] ,
\end {equation}
where $f(z)$ is a continuous, non-singular function of $z$,
corresponding to $z {\cal F}(x,y)$ in our case.
{\it If} the bounds on integration over $z$ were symmetric about $z{=}0$,
we would be able to average the integrand with $z\to-z$ to write
\begin {equation}
  \int_{-a}^a dz \> f(z) \, \PV\Bigl[ \frac{1}{z} \Bigr]
  =
  \int_{-a}^a dz \> \frac{f(z) - f(-z)}{2 z}
  \,.
\label {eq:trick}
\end {equation}
The integrand on the right-hand side is finite at $z{=}0$, and so
it (i) no
longer needs the PV prescription and (ii) is suitable for numerical
integration.
Unfortunately, our actual integration interval is {\it not}\/ symmetric under
$z {\to} -z$. We must divide the integration into different
integration regions (one symmetric around $z{=}0$ and another that
avoids $z{=}0$) and treat them differently.
The shaded region of fig.\ \ref{fig:PV} shows the largest region of
$y$ that is symmetric under
$y \to 2(1{-}x) - y$, which is the transformation that
takes $z \to -{z}$ without changing $x$.
Using (\ref{eq:trick}) for the shaded region,
the integral in (\ref{eq:classII}) can then be rewritten as
the numerics-friendly expression
\begin {equation}
  \left[ \frac{d\Gamma}{dx} \right]_{\rm F,~virt~II}
  =
  2\Re
  \int_0^1 dy
  \begin {cases}
     \tfrac12 \bigl[ {\cal F}(x,y) + {\cal F}\bigl(x,2(1{-}x){-}y\bigr) \bigr];
         & 1{-}2x \le y \le 2{-}2x , \\
     {\cal F}(x,y) & \mbox{otherwise} .
  \end {cases}
\label {eq:average}
\end {equation}

% ============================================================================

\section {Conclusion}
\label {sec:conclusion}

A summary of formulas for
the final results of this paper is given in appendix
\ref{app:summary}.  It is natural to wonder how much
quantitative impact the processes of this paper
(figs.\ \ref{fig:FdiagsReal}--\ref{fig:FdiagsVirtII})
will have compared to the ``nearly-complete'' calculation
of ref.\ \cite{qcd}.
With the tools presented so far, this is a somewhat ambiguous question.
For example, virtual diagrams must be integrated over $y$.
But the same integration of the virtual diagrams of ref.\ \cite{qcd}
gives an infrared divergence, which cannot be meaningfully
compared to the non-divergent results of this paper.%
\footnote{
  One might try comparing the size of the $y$-integrands, but this is also
  meaningless for virtual diagrams.
  We know from the discussion in section \ref{sec:PV}
  of the original $y$-integrand of (\ref{eq:calF}) that the original
  integrand is $\pm\infty$ at $y = 1{-}x$.  That can't be meaningfully
  compared to the size of another diagram's $y$-integrand because the
  divergence goes away when integrated.  One might look at the integrand of
  (\ref{eq:average}) instead, but the details of that integrand depend
  on our arbitrary choice of exactly which region to apply $z{\to}-z$
  averaging to.  Only the integral over $y$ has meaning there, not the
  $y$-integrand by itself.
}
Even if one adds together the real and virtual diagrams of ref.\ \cite{qcd},
there is still a double-log infrared divergence.
A later paper \cite{finale}
will discuss how to make infrared-safe calculations of in-medium
shower development, for
which the relative size of contributions can then be examined.
We leave that comparison until then.

%%%%%%%%%%%%%%%%%%%%%%%%%%%%%%%%%%%%%%%%%%%%%%%%%%%%%%%%%%%%%%%%%%%%%%%%%%%%%%

\acknowledgments

We are indebted to Yuri Kovchegov and Risto Paatelainen for
conversations about regularization of the $q^+{=}0$ divergences
of longitudinal gluon exchange in lightcone perturbation theory.
This work was supported, in part, by the U.S. Department
of Energy under Grants No.~DE-SC0007984 and DE-SC0007974 (Arnold);
by the Deutsche Forschungsgemeinschaft
(DFG, German Research Foundation) -- Project-ID 279384907 -- SFB 1245
and by the State of Hesse within the Research Cluster ELEMENTS
(Project ID 500/10.006) (Gorda);
and by the National Natural
Science Foundation of China under
Grant Nos.\ 11935007, 11221504 and 11890714 (Iqbal).

%%\paragraph{Note added.} This is also a good position for notes added
%%after the paper has been written.

%%%%%%%%%%%%%%%%%%%%%%%%%%%%%%%%%%%%%%%%%%%%%%%%%%%%%%%%%%%%%%%%%%%%%%%%%%%%%%
\appendix
%%%%%%%%%%%%%%%%%%%%%%%%%%%%%%%%%%%%%%%%%%%%%%%%%%%%%%%%%%%%%%%%%%%%%%%%%%%%%%

\section{Summary}
\label{app:summary}

Appendix A of ref.\ \cite{qcd} gave a summary, for the ``nearly-complete''
calculation there, of all rates associated with overlap effects in
sequential gluon splitting.  Here, we summarize how to add in
the remaining diagrams analyzed in this paper.

% ---------------------------------------------------------------------------

\subsection{$g\to ggg$ rate}
\label {app:gggSummary}

Eq.\ (A.9) of ref.\ \cite{qcd} for the total overlap effect on
real double splitting should be modified to
\begin {equation}
   \left[ \Delta \frac{d\Gamma}{dx\,dy} \right]_{g{\to}ggg}
   =
   \left[ \frac{d\Gamma}{dx\,dy} \right]_{\rm crossed}
   +
   \left[ \Delta \frac{d\Gamma}{dx\,dy} \right]_{\rm seq}
   +
   \left[ \frac{d\Gamma}{dx\,dy} \right]_{{\rm F}} ,
\end {equation}
where the new term is
\begin {equation}
   \left[ \frac{d\Gamma}{dx\,dy} \right]_{{\rm F}}
   =
   \left[ \frac{d\Gamma}{dx\,dy} \right]_{{\rm single~F}}
   +
   \left[ \frac{d\Gamma}{dx\,dy} \right]_{({\rm FF})} .
\end {equation}

% .......................................................................

\subsubsection {single F piece}

The ``single F'' piece corresponds to the analogous 4-gluon vertex result
of section 4.1 of ref.\ \cite{4point} but with the substitution
$\zeta_{(4)} \rightarrow \zeta_{\rm(F)}$ derived in this paper.  That
has the form
\begin {align}
   \left[ \frac{d\Gamma}{dx\>dy} \right]_{{\rm single~F}}
   = \quad
   & {\cal A}_{\rm(F)}(x,y) + {\cal A}_{\rm(F)}(1{-}x{-}y,y)
                        + {\cal A}_{\rm(F)}(x,1{-}x{-}y)
\nonumber\\
   + ~ &
   {\cal A}_{\rm(F)}(y,x) + {\cal A}_{\rm(F)}(y,1{-}x{-}y)
                      + {\cal A}_{\rm(F)}(1{-}x{-}y,x) ,
\label {eq:dGamma4}
\end {align}
where ${\cal A}_{\rm(F)}(x,y)$ is
the result of one color routing of
$F\bar y\bar x + \bar y F\bar x + \bar y \bar x F$
(from fig.\ \ref{fig:FdiagsReal}) plus conjugates.
We'll find it convenient later, for evaluating virtual diagrams,
to split ${\cal A}_{\rm(F)}$ into separate contributions from each non-zero
diagram (plus its conjugate):
\begin {equation}
   {\cal A}_{\rm(F)}(x,y)
   \equiv
   {\cal A}_{F\bar y\bar x}(x,y) + {\cal A}_{\bar y\bar x F}(x,y) ,
\end {equation}
where
\begin {subequations}
\begin {align}
   {\cal A}_{F\bar y\bar x}(x,y)
   &\equiv
   \int_0^{+\infty} d(\Delta t) \>
        2 \Re \bigl( B_{F\bar y\bar x}(x,y,\Delta t) \bigr) ,
\\
   {\cal A}_{\bar y\bar x F}(x,y)
   &\equiv
   \int_0^{+\infty} d(\Delta t) \>
        2 \Re \bigl( B_{\bar y\bar x F}(x,y,\Delta t) \bigr) ,
\end {align}
\end {subequations}
\begin {subequations}
\label {eq:BFsummary}
\begin {align}
   B_{F\bar y\bar x}(x,y,\Delta t) &=
       D_{\rm(F)}(\hat x_1,\hat x_2,\hat x_3,\hat x_4,\zeta_{\rm(F)},\Delta t)
       = D_{\rm(F)}({-}1,y,1{-}x{-}y,x,\zeta_{\rm(F)},\Delta t) ,
\\
   B_{\bar y\bar x F}(x,y,\Delta t) &=
       D_{\rm(F)}(-\hat x_4,-\hat x_3,-\hat x_2,-\hat x_1,\zeta_{\rm(F)},\Delta t)
       = D_{\rm(F)}(-x,-(1{-}x{-}y),-y,1,\zeta_{\rm(F)},\Delta t) ,
\end {align}
\end {subequations}
where $\zeta_{\rm(F)} \equiv \zeta_{\rm(F)}(x,y)$.
Here, we follow the notation of
Appendix A of ref.\ \cite{qcd} by using hats over
$(x_1,x_2,x_3,x_4)$ to represent our usual numbering convention
(\ref{eq:x1234}):
\begin {equation}
  (\hat x_1,\hat x_2,\hat x_3,\hat x_4) \equiv (-1,y,1{-}x{-}y,x) .
\end {equation}
Below, the $(x_1,x_2,x_3,x_4)$ without hats
will instead generically
represent whatever the arguments of the function $D$ are.
\begin {align}
   D_{\rm(F)}(x_1,&x_2,x_3,x_4,\zeta,\Delta t) =
\nonumber\\ &
   - \frac{\CA^2\alphas^2 M_\fx}{16 \pi^2 E} \,
   (-x_1 x_2 x_3 x_4)
   \,\zeta\,
   \Omega_+ \Omega_- \csc(\Omega_+\,\Delta t) \csc(\Omega_-\,\Delta t)
   \frac{Y_\ybx}{X_\ybx} \,,
\label {eq:DF}
\end {align}
where the low-level expressions for the symbols
$M_\fx$, $\Omega_\pm$, $X_\ybx$ and $Y_\ybx$ in terms of
the arguments $(x_1,x_2,x_3,x_4,\Delta t)$ are the same as in
appendices A.2.1 and A.2.2 of ref.\ \cite{qcd}.

In (\ref{eq:BFsummary}), the argument $\zeta_{\rm(F)}$ passed to $D_{\rm(F)}$ is
\begin {equation}
  \zeta_{\rm(F)}
  =
   \zeta_{(4)}
   + i_{12} \, \zeta_{12}
   - i_{14} \, \zeta_{14} ,
\label {eq:zetaFapx}
\end {equation}
where
\begin {subequations}
\begin {equation}
   \zeta_{12} =
   \frac{(x^2{+}z^2)(1{+}y^2)}
        {(x y z)^2(1{-}y)^3} \,,
   \qquad
   \zeta_{13} =
   \frac{(1{-}y)^4 + z^2 + x^2 y^2}
        {(x y z)^2(1{-}y)^3} \,,
   \qquad
   \zeta_{14} =
   \frac{(1{-}y)^4 + x^2 + z^2 y^2}
        {(x y z)^2(1{-}y)^3} \,,
\end {equation}
\begin {equation}
   \zeta_{(4)} = \zeta_{12} - 2\zeta_{13} + \zeta_{14} ,
\end {equation}
\begin {equation}
   i_{12} = \frac{(1{+}y)(x{-}z)}{(1{-}y)^2} \,,
   \qquad
   i_{14} = \frac{(1{+}x)(z{-}y)}{(1{-}x)^2} \,.
\end {equation}
\end {subequations}

% .......................................................................

\subsubsection {FF piece}

The FF piece corresponds to the $F\bar F$ diagram of fig.\ \ref{fig:FdiagsReal}
plus its complex conjugate (which corresponds to the other time ordering,
$\bar F F$).  For the canonical color routing, the FF piece is given by
(i) the 4-gluon vertex result for ${\cal A}_{(44)}$ in
section 4.2 of ref.\ \cite{4point} times (ii) the factor indicated
in our (\ref{eq:convert44}).
The sum over color routings is then
\begin {equation}
   \left[ \frac{d\Gamma}{dx\>dy} \right]_{\rm(FF)}
   =
   {\cal A}_{\rm(FF)}(x,y) + {\cal A}_{\rm(FF)}(1{-}x{-}y,y)
                        + {\cal A}_{\rm(FF)}(x,1{-}x{-}y)
\label {eq:dGamma44}
\end {equation}
with
\begin {equation}
   {\cal A}_{\rm(FF)}(x,y)
   \equiv
   \bigl[
     1 +
     \tfrac13 ( i_{12}^2 + i_{14}^2 - i_{12} i_{14} + i_{12} - i_{14} )
   \bigr]
   {\cal A}_{(44)}(x,y)
\end{equation}
and
\begin {equation}
   {\cal A}_{(44)}(x,y)
   \equiv
   \int_0^{+\infty} d(\Delta t) \>
        2 \Re \bigl( B_{(44)}(x,y,\Delta t) \bigr) ,
\end {equation}
\begin {align}
   B_{(44)}(x,y,\Delta t) &=
       C_{(44)}(\hat x_1,\hat x_2,\hat x_3,\hat x_4,\Delta t)
   =
       C_{(44)}({-}1,y,1{-}x{-}y,x,\Delta t) ,
\end {align}
\begin {equation}
   C_{(44)} = D_{(44)} - \lim_{\hat q\to 0} D_{(44)} ,
\end {equation}
\begin {equation}
   D_{(44)}(x_1,x_2,x_3,x_4,\Delta t) =
   - \frac{3\CA^2 \alphas^2}{16 \pi^2} \,
   \Omega_+\Omega_- \csc(\Omega_+\,\Delta t) \csc(\Omega_-\,\Delta t)
   .
\end {equation}

% ---------------------------------------------------------------------------

\subsection{NLO $g\to gg$ rate}

The virtual corrections to single splitting $g {\to} gg$ are
written in Appendix A.3 of ref.\ \cite{qcd} in terms of
\begin {align}
   \left[ \Delta \frac{d\Gamma}{dx} \right]^{\NLObar}_{g\to gg}
   &=
     \biggl(
       \left[ \Delta \frac{d\Gamma}{dx} \right]_\virtI
     \biggr)
     + (x \to 1{-}x)
   + \left[ \Delta \frac{d\Gamma}{dx} \right]_\virtII
\nonumber\\
   &=
     \biggl(
       \int_0^{1-x} dy \, \left[ \Delta \frac{d\Gamma}{dx\,dy} \right]_\virtI
     \biggr)
     + (x \to 1{-}x)
   +
   \int_0^1 dy \, \left[ \Delta \frac{d\Gamma}{dx\,dy} \right]_\virtII ,
\label {eq:dGammaNLObar}
\end {align}
where the three terms are, in order, the contribution of class I diagrams,
their $x{\to}1{-}x$ cousins, and class II diagrams.
Eqs.\ (A.53) and (A.54) of ref.\ \cite{qcd} for the Class I and II
$y$-integrands should be modified to
\begin {equation}
  \left[ \Delta \frac{d\Gamma}{dx\,dy} \right]_\virtI
  =
  \left[ \frac{d\Gamma}{dx\,dy} \right]_\virtIc
  +
  \left[ \Delta \frac{d\Gamma}{dx\,dy} \right]_\virtIs
  +
  2\Re \left[ \frac{d\Gamma}{dx\,dy} \right]_{xyy\bar x}
  + \left[ \frac{d\Gamma}{dx\,dy} \right]_\virtIf
\label {eq:virtIsummary}
\end {equation}
and
\begin {equation}
  \left[ \Delta \frac{d\Gamma}{dx\,dy} \right]_\virtII
  =
  \left[ \Delta \frac{d\Gamma}{dx\,dy} \right]_\virtIIs
  +
  2\Re \left[ \frac{d\Gamma}{dx\,dy} \right]_{x\bar y\bar y\bar x}
  + \left[ \frac{d\Gamma}{dx\,dy} \right]_\virtIIf
  ,
\end {equation}
where the new addition is the last term in each.

% ...........................................................................

\subsubsection {$[d\Gamma/dx\, dy]_\virtIf$}

Given that $[d\Gamma/dx\,dy]_\virtIf$ will be integrated over
$0<y<1{-}x$ in (\ref{eq:dGammaNLObar}),
there are two equivalent ways to choose the integrand.  One way is to
include both color routings of the diagrams for every value of
$y$ (the routings are related by $y \to 1{-}x{-}y$) and write the
$y$-integrand in the form
\begin {subequations}
\label {eq:virtIf}
\begin {equation}
   \left[ \frac{d\Gamma}{dx\,dy} \right]_\virtIf
   = \tfrac12 \bigl[
       {\cal A}_\virtIf(x,y) + {\cal A}_\virtIf(x,1{-}x{-}y)
     \bigr] ,
\label {eq:AvertIf1}
\end {equation}
where $\tfrac12$ is the loop symmetry factor associated with the
diagrams and ${\cal A}_\virtIf$ is the result for a single color routing
{\it without} including any loop symmetry factor.
But, because of the $y \leftrightarrow 1{-}x{-}y$ symmetry of
(\ref{eq:AvertIf1}),
the $y$ integral is the same if we integrate only one color routing but
drop the loop symmetry factor, and so instead take
\begin {equation}
   \left[ \frac{d\Gamma}{dx\,dy} \right]_\virtIf
   = {\cal A}_\virtIf(x,y) .
\end {equation}
\end {subequations}
Either way, (\ref{eq:FXby}) and (\ref{eq:FyXb}) give
\begin {align}
  {\cal A}_\virtIf(x,y)
  &=
  -{\cal A}_{F\bar y\bar x}(y,x)
  +\Bigl[
    {\cal A}_{F\bar y\bar x}(\tfrac{-x}{1-x-y} , \tfrac{-y}{1-x-y} )
   \Bigr]_{E \to (1-x-y)E}
\nonumber\\
  &=
  -{\cal A}_{F\bar y\bar x}(y,x)
  +(1{-}x{-}y)^{-1/2}
    {\cal A}_{F\bar y\bar x}(\tfrac{-x}{1-x-y} , \tfrac{-y}{1-x-y} ) ,
\end {align}
where the last line follows from the fact that rates $d\Gamma/dx\,dy$
are proportional to $\sqrt{\qhatA/E}$.%
\footnote{
  See the discussion of similar examples of this scaling argument
  in appendices D.3 and D.5 of ref.\ \cite{qcd}.
}

% ...........................................................................

\subsubsection {$[d\Gamma/dx\, dy]_\virtIIf$}

Class II diagrams are integrated over $0 < y < 1$ in (\ref{eq:dGammaNLObar}).
Analogous to (\ref{eq:virtIf}), we may write
\begin {subequations}
\label {eq:virtIIf}
\begin {equation}
   \left[ \frac{d\Gamma}{dx\,dy} \right]_\virtIIf
   = \tfrac12 \bigl[
       \,\overline{\cal A}_\virtIIf(x,y) + \overline{\cal A}_\virtIIf(x,1{-}y)
     \bigr]
\end {equation}
or
\begin {equation}
   \left[ \frac{d\Gamma}{dx\,dy} \right]_\virtIIf
   = \overline{\cal A}_\virtIIf(x,y) .
\end {equation}
\end {subequations}
The latter is equivalent to the version presented in
(\ref{eq:classIIcanon}), with ${\cal A}_\virtIIf$ here representing
$2\Re{\cal F}$.
The overlines on ${\cal A}_\virtIIf$ in (\ref{eq:virtIIf}) will represent
the averaging procedure of section \ref{sec:average}.  If we
were instead content with $y$-integrands that had divergences
requiring implementation of a principal value prescription, we could
drop the overlines, and
eqs.\ (\ref{eq:yXbF}) and (\ref{eq:XbyF}) give
\begin {subequations}
\begin {align}
  {\cal A}_\virtIIf(x,y) =
  &=
  -\Bigl[
    {\cal A}_{\bar y\bar x F}(\tfrac{x}{1-y},\tfrac{-y}{1-y})
   \Bigr]_{E \to (1-y)E}
  +\Bigl[
    {\cal A}_{\bar y\bar x F}(\tfrac{y}{1-x} , \tfrac{-x}{1-x} )
   \Bigr]_{E \to (1-x)E}
\nonumber\\
  &=
  -(1{-}y)^{-1/2}
    {\cal A}_{\bar y\bar x F}(\tfrac{x}{1-y},\tfrac{-y}{1-y})
  +(1{-}x)^{-1/2}
    {\cal A}_{\bar y\bar x F}(\tfrac{y}{1-x} , \tfrac{-x}{1-x} )
  .
\end {align}
\end {subequations}
Following (\ref{eq:average}), our numerics-friendly, averaged version
$\overline{\cal A}_\virtIIf$ of
${\cal A}_\virtIIf$ is
\begin {equation}
  \overline{\cal A}_\virtIIf(x,y)
  \equiv
  \begin {cases}
     \tfrac12 \bigl[
        {\cal A}_\virtIIf(x,y) + {\cal A}_{\virtIIf}\bigl(x,2(1{-}x){-}y\bigr)
     \bigr];
         & 1{-}2x \le y \le 2{-}2x , \\
     {\cal A}_\virtIIf(x,y) & \mbox{otherwise} .
  \end {cases}
\end {equation}

%%%%%%%%%%%%%%%%%%%%%%%%%%%%%%%%%%%%%%%%%%%%%%%%%%%%%%%%%%%%%%%%%%%%%%%%%%%%%%

\end{document}